\documentclass[aps,onecolumn,superscriptaddress,prl,floatfix,british]{revtex4}

\usepackage{amssymb}
\usepackage{amsmath}
\usepackage{babel}
\usepackage{graphicx}
\usepackage{epstopdf}
\usepackage[sort&compress]{natbib}
\usepackage{topcapt}
\usepackage{hyperref}
\usepackage{xcolor}

\begin{document}
\title{Quantum simulation of photosynthetic energy transfer}

\author{Bi-Xue Wang}
\thanks{These authors contributed equally to this work.}
\affiliation{State Key Lab for Low-dimensional Quantum Physics and Department of Physics, Tsinghua University, Beijing 100084}

\author{Ming-Jie Tao}
\thanks{These authors contributed equally to this work.}
\affiliation{Department of Physics, Applied Optics Beijing Area Major Laboratory,
Beijing Normal University, Beijing 100875}
\affiliation{State Key Lab for Low-dimensional Quantum Physics and Department of Physics, Tsinghua University, Beijing 100084}

\author{Qing Ai}
\thanks{These authors contributed equally to this work.}
\affiliation{CEMS, RIKEN, Wako-shi, Saitama 351-0198, Japan}
\affiliation{Department of Physics, Applied Optics Beijing Area Major Laboratory,
Beijing Normal University, Beijing 100875}

\author{Tao Xin}
\affiliation{State Key Lab for Low-dimensional Quantum Physics and Department of Physics, Tsinghua University, Beijing 100084}

\author{Neill Lambert}
\affiliation{CEMS, RIKEN, Wako-shi, Saitama 351-0198, Japan}

\author{Dong Ruan}
\affiliation{State Key Lab for Low-dimensional Quantum Physics and Department of Physics, Tsinghua University, Beijing 100084}

\author{Yuan-Chung Cheng}
\affiliation{Department of Chemistry, National Taiwan University, Taipei  City 106}

\author{Franco Nori}
\affiliation{CEMS, RIKEN, Wako-shi, Saitama 351-0198, Japan}
\affiliation{Physics Department, The University of Michigan, Ann Arbor, Michigan 48109-1040, USA}

\author{Fu-Guo Deng}
\affiliation{Department of Physics, Applied Optics Beijing Area Major Laboratory,
Beijing Normal University, Beijing 100875}
\affiliation{NAAM-Research Group, Department of Mathematics, Faculty of Science,
King Abdulaziz University, Jeddah 21589, Saudi Arabia}

\author{Gui-Lu Long}
\affiliation{State Key Lab for Low-dimensional Quantum Physics and Department of Physics, Tsinghua University, Beijing 100084}
\affiliation{Beijing National Research Center on Information Science and Technology, Beijing 100084}
\affiliation{School of Information Science and Technology, Tsinghua University, Beijing 100084}
\affiliation{Innovation Center of Quantum Matter, Beijing 100084}

\date{\today}
\maketitle

\textbf{Near-unity energy transfer efficiency has been widely
observed in natural photosynthetic complexes.
This phenomenon has attracted broad interest from different fields,
such as physics, biology, chemistry and material science,
as it may offer valuable insights into efficient solar-energy harvesting.
Recently, quantum coherent effects have been discovered in
photosynthetic light harvesting,
and their potential role on energy transfer
has seen heated debate.
Here, we perform an experimental quantum simulation of
photosynthetic energy transfer using nuclear magnetic resonance (NMR).
We show that an $N$-chromophore photosynthetic complex,
with arbitrary structure and bath spectral density,
can be effectively simulated by a system with $\log_{2}N$ qubits.
The computational cost of simulating such a system with a theoretical tool,
like the hierarchical equation of motion, which is exponential in $N$,
can be potentially reduced to requiring a just polynomial number of qubits $N$
using NMR quantum simulation.
The benefits of performing such quantum simulation in NMR are even greater
when the spectral density is complex,
as in natural photosynthetic complexes.
These findings may shed light on quantum coherence
in energy transfer and help to provide design principles for efficient
artificial light harvesting.}


Efficient exciton energy transfer (EET) is crucial in photosynthesis and solar cells \cite{Cheng09,Romero17},
especially when the systems are large \cite{Scholes17},
e.g., as in Photosystem I (PSI) and Photosystem II (PSII),
which have hundreds of chromophores.
A comprehensive knowledge of the quantum dynamics of such systems
would be of potential importance to the study on EET \cite{Lambert13}.
Much effort has been made to reveal the effects of quantum coherence
on efficient energy transfer \cite{Lambert13,Engel07,Chin13,Collini10,Hildner13,Dong12,Ishizaki09}.
In order to try to mimic EET,
a Frenkel-exciton Hamiltonian is required and
this can be studied with quantum chemistry approaches \cite{Cheng09},
e.g., fitting experimental spectra, or calculations by density functional theory.
Because photosynthetic pigment-protein complexes are intrinsically open quantum systems,
with system-bath couplings comparable to the intra-system couplings,
it is difficult to faithfully mimic the exact quantum dynamics of EET.
Among the methods for describing EET \cite{Cheng09,Ishizaki09,Ai13},
the hierarchical equation of motion (HEOM) yields a numerically exact solution
at the cost of considerable computing time \cite{Tanimura06,Ishizaki09}. For the widely-studied Fenna-Matthews-Olson (FMO) complex, a reliable result could be produced by 16,170 coupled differential equations at low temperatures, based on the HEOM \cite{Tanimura06,Ishizaki09} with 4 layers. Moreover, the HEOM encounters difficulties when the system
has non-trivial spectral density,
making it often difficult to verify the results.
On the other hand, due to heterogeneity, e.g. static disorder and conformational change, it is
difficult to experimentally verify the theoretical predictions in natural
photosynthetic systems. Because quantum simulations with $N$ qubits can powerfully mimic
the quantum dynamics of $2^N$ states by virtue of quantum mechanics \cite{Buluta09,Georgescu14,Buluta11},
the quantum simulation of photosynthetic energy transfer with an arbitrary Hamiltonian
by nuclear magnetic resonance (NMR) provides an intriguing approach to verify theoretical predictions. Recently, a newly-developed technique which could simulate
the effect of a bath  with an arbitrary spectral density
by a set of classical pulses has been successfully realized with ion traps \cite{Soare14} and NMR \cite{Zhen16}. Therefore, photosynthetic light-harvesting with an arbitrary Hamiltonian
and spectral density,
describing a structured environment, can be experimentally simulated by NMR.

\begin{figure}[htbp]
\centering
\includegraphics[width=85mm,angle=0]{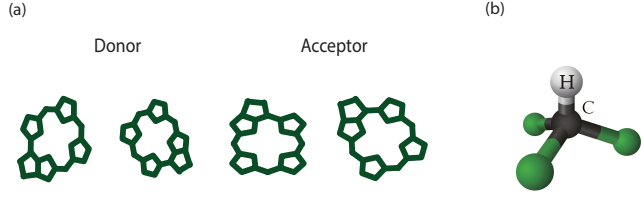}
~\caption{{\bfseries Photosynthetic chromophore arrangement and physical system for NMR simulation.} (a) Linear geometry with four chromophores for photosynthetic energy transfer; (b) Chemical structure for a $^{13}$C-labeled chloroform molecule, where the H and C nuclear spins are chosen as the two qubits.
\label{fig:scheme}}
\end{figure}

NMR is an excellent platform for quantum simulation since it is easy
to operate and it can have long coherence times \cite{Luo17,Feng13}.
In this paper, NMR is utilized to simulate the quantum coherent dynamics in photosynthetic
light harvesting. As a prototype, a tetramer including four chlorophylls \cite{Ai13},
as schematically illustrated in Fig.~\ref{fig:scheme}(a), is
employed in the NMR quantum simulation. In a previous investigation \cite{Ai13},
the tetramer model was exploited to study the clustered geometry utilizing
exciton delocalization and energy matching to accelerate the energy transfer.
The EET in photosynthesis is described
by the Frenkel-exciton Hamiltonian
\begin{eqnarray}
H_{\textrm{EET}}&=&\sum_{i=1}^4\varepsilon_i\vert i\rangle\langle i\vert+\sum_{i\neq j=1}^4 J_{ij}\vert i\rangle\langle j\vert,
\label{eq:Heff}
\end{eqnarray}
where $\vert i\rangle$ ($i=1,2,3,4$) is the state with a single excitation at site $i$
and other sites at the ground state, $\varepsilon_i$ is the site energy of $\vert i\rangle$,
$J_{ij}$ is the excitonic interaction between sites $i$ and $j$.
In our quantum simulations, we adopt the site energies \cite{Ai13} $\varepsilon_1=13000$ cm$^{-1}$, $\varepsilon_2=12900$ cm$^{-1}$, $\varepsilon_3=12300$ cm$^{-1}$, and $\varepsilon_4=12200$ cm$^{-1}$, and the couplings $J_{12}=J_{34}=126$ cm$^{-1}$, $J_{13}=J_{24}=16$ cm$^{-1}$, $J_{23}=132$ cm$^{-1}$, and $J_{14}=5$ cm$^{-1}$, which are typical parameters in
photosynthetic systems.

For a photosynthetic complex with $N$ chlorophylls,
there are only $N$ single-excitation states involved in the energy transfer. Therefore, only $\log_{2}N$ qubits are required
to realize the quantum simulation. To mimic the energy transfer described by
the above Hamiltonian (\ref{eq:Heff}), two qubits are necessary for the quantum simulation.
In this case, the photosynthetic single-excitation state $\vert i\rangle$ is encoded as a two-qubit product state,
i.e. $\vert 00\rangle$, $\vert 01\rangle$,
$\vert 10\rangle$, and $\vert 11\rangle$.
By a straightforward calculation, the Frenkel-exciton Hamiltonian can
be mapped to the NMR Hamiltonian as
\begin{eqnarray}
H_{\textrm{NMR}}\!\!\!\!\!\!&&=\frac{\varepsilon_1^\prime+\varepsilon_2^\prime-\varepsilon_3^\prime-\varepsilon_4^\prime}{4}\sigma_1^z
+\frac{\varepsilon_1^\prime-\varepsilon_2^\prime+\varepsilon_3^\prime-\varepsilon_4^\prime}{4}\sigma_2^z \notag \\
&&+\frac{J_{13}^\prime+J_{24}^\prime}{2}\sigma_1^x
+\frac{J_{12}^\prime+J_{34}^\prime}{2}\sigma_2^x
+\frac{J_{14}^\prime+J_{23}^\prime}{2}\sigma_1^x\sigma_2^x \notag \\
&&+\frac{J_{23}^\prime-J_{14}^\prime}{2}\sigma_1^y\sigma_2^y
+\frac{J_{13}^\prime-J_{24}^\prime}{2}\sigma_1^x\sigma_2^z \notag \\
&&+\frac{J_{12}^\prime-J_{34}^\prime}{2}\sigma_1^z\sigma_2^x,
\label{eq:Hnmr}
\end{eqnarray}
where $\sigma_j^u$ ($j=1,2$, $u=x,y,z$) is the Pauli operator for qubit $j$, numerically $\varepsilon_j^\prime=\pi\varepsilon_j/10$ and $J_{ij}^\prime=\pi J_{ij}/10$,
but the dimension cm$^{-1}$ should be replaced by kHz. In other words,
all realistic parameters have been scaled down in energy by a factor of $1~\textrm{cm}^{-1}/(\pi/10~\textrm{kHz})=3\times10^8/\pi$.

The excitonic coupling $J_{ij}$ between sites $i$ and $j$
makes the exciton energy hop in both directions,
i.e. $\vert i\rangle\leftrightarrows\vert j\rangle$.
Apart from this, the system-bath couplings will
facilitate the energy flow towards an energy trap,
where the captured photon energy is converted into
chemical energy \cite{Romero17,Scholes17}.
The interaction between the system and bath in photosynthetic complexes
can be described by
\begin{eqnarray}
H_{\textrm{SB}}=\sum_{i,k}g_{ik}\vert i\rangle\langle
i\vert(a_{ik}^\dag+a_{ik}),\label{eq:Hsb}
\end{eqnarray}
where $a_{ik}^\dag$ is the creation operator for the $k$th phonon mode
of chlorophyll $i$ with coupling strength $g_{ik}$.
$H_{\textrm{SB}}$ will induce pure-dephasing on the $i$-th chromophore
when it is in the excited state. Generally,
the system-bath couplings are given by the spectral density
\begin{eqnarray}
G_{\textrm{EET}}(\omega)=\sum_{k}g_{ik}^2\delta(\omega-\omega_k),
\end{eqnarray}
which we assume identical for all chromophores.
For typical photosynthetic complexes, the system-bath couplings are of the same
order as the intra-system couplings.
In order to mimic the effects of noise, we utilize the bath-engineering technique
which has been successfully implemented in ion traps and NMR.
The implementation of a pure-dephasing Hamiltonian
\begin{equation}
H_{\textrm{PDN}}=\vec{B}(t)\cdot\vec{\sigma} \label{eq:Hpdn}
\end{equation}
relies on generating stochastic errors by
performing phase modulations on a constant-amplitude carrier, i.e.
\begin{eqnarray}
\vec{B}(t)&=&\Omega_0\cos[\omega_{\mu}t+\phi_N(t)]\hat{z}, \label{eq:Bz}\\
\phi_N(t)&=&\alpha\sum_{j=1}^{J}F(j)\sin(\omega_{j}t+\psi_j), \label{eq:phi}
\end{eqnarray}
where $\Omega_0$ is the constant amplitude of a magnetic field with driving frequency $\omega_{\mu}$,
$\psi_j$ is a random number, $\omega_{j}=j\omega_{0}$ with $\omega_0$ and $\omega_J=J\omega_0$ being base and cutoff frequencies respectively, and $\alpha$ is a global scaling factor.
The power-spectral density $S_z(\omega)\equiv\int d\tau\langle\dot{\phi}_N(t+\tau)\dot{\phi}_N(t)\rangle e^{i\omega\tau}$ is the Fourier transform of the autocorrelation function
of $\dot{\phi}_N(t)$,
\begin{eqnarray}
S_z(\omega)\!\!\!&=&\!\!\!\frac{\pi\alpha^2\omega_0^2}{2}\sum_{j=1}^{J}j^2F^2(j)[\delta(\omega-\omega_j)+\delta(\omega+\omega_j)].
\end{eqnarray}
By appropriately choosing $F(j)$ and the cutoff $J$, an arbitrary power-spectral density of the bath can be realized, e.g. white, Ohmic, or Debye spectral densities.
For the details, please refer to the Supplementary Information.

As illustrated in Fig.~\ref{fig:scheme}(b), the nuclear spins of the carbon atom and
hydrogen atom in a chloroform molecule are chosen to encode the two
qubits with the Hamiltonian written as
\begin{eqnarray}
H_{\textrm{CHCl}_3}&=& \pi\omega_1\sigma_1^z+\pi\omega_2\sigma_2^z+\frac{\pi J}{2}\sigma_1^z\sigma_2^z,
\label{eq:CHCl3}
\end{eqnarray}
where $\omega_1=3206.5$ Hz, $\omega_2=7787.9$ Hz are the chemical shifts of the two spins,
and $J= 215.1$ Hz \cite{Chuang98}.
Therefore, in order to simulate the quantum dynamics of energy
transfer, the entire evolution is decomposed into repetitive identical cycles.
After each cycle, there is a small difference between the exact evolution
and the simulated one. However, because there are tens of thousands of cycles before
completing the energy transfer, the accumulated error might be significant so
that the simulation becomes unreliable. To avoid this problem,
we utilize the gradient ascent pulse engineering (GRAPE) algorithm~\cite{Khaneja05}, which has been successfully applied to a number of quantum simulations in NMR~\cite{Zhang11,Feng13,Peng14},
to mimic the quantum dynamics of light harvesting.

\begin{figure}[htbp]
\begin{center}
\includegraphics[width=160mm]{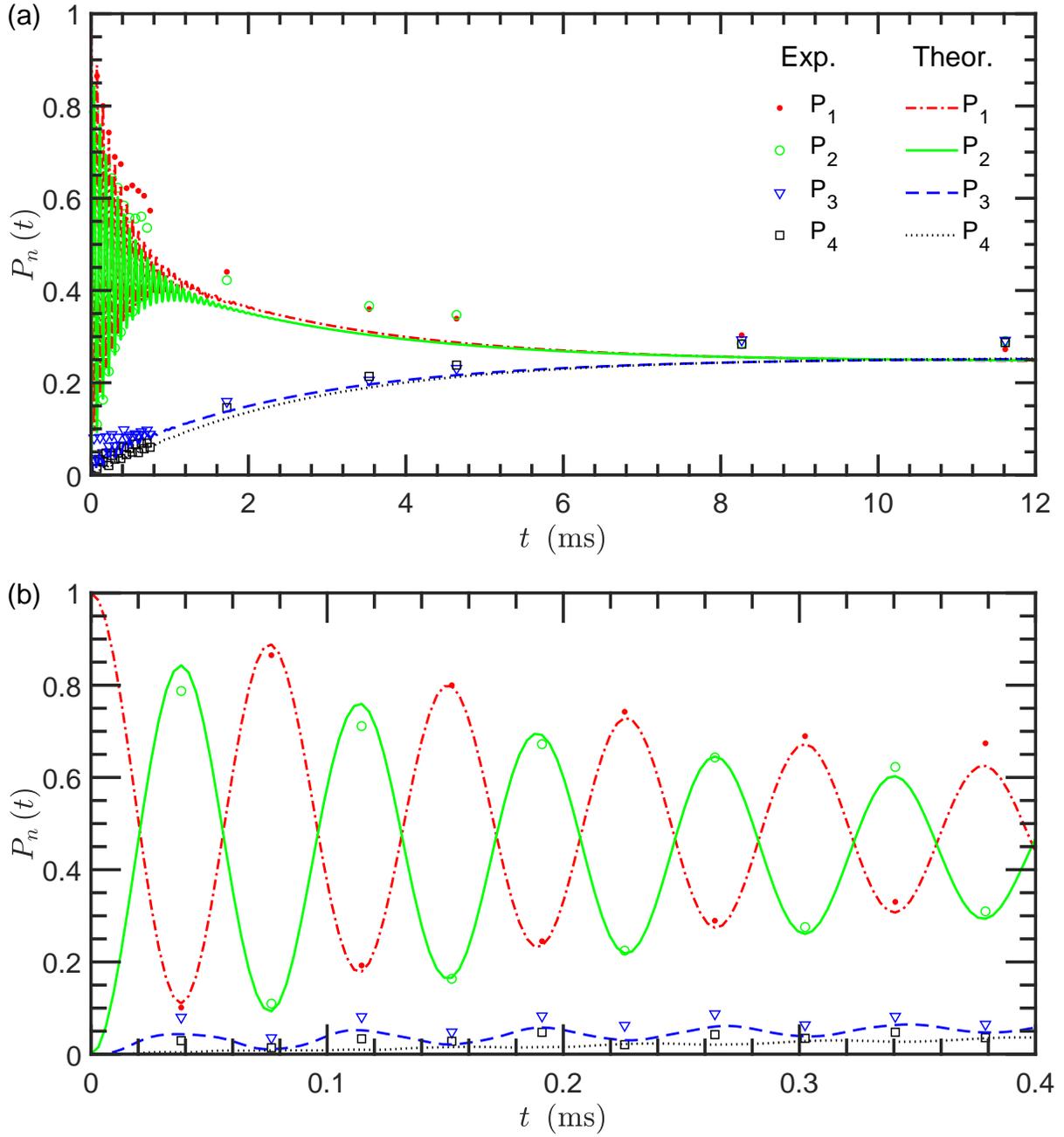}%
\caption{{\bfseries Simulation of the energy transfer governed by
$H_{\textrm{NMR}}+H_{\textrm{PDN}}$ for $M=150$ random realizations.}  (a) Long-time quantum dynamics; (b) short-time quantum dynamics. The dots show the experimental data, and the curves are obtained from the numerical simulation using the HEOM. In all figures, the horizontal ordinates of the curves for EET dynamics have been magnified by $3\times10^8/\pi$ times.\label{fig:Dynamics}}
\end{center}
\end{figure}

In Fig.~\ref{fig:Dynamics}(a), the quantum coherent energy transfer
for the above Hamiltonian $H_\textrm{NMR}+H_\textrm{PDN}$,
using an initial condition where an excitation is localized on the first chromophore,
is simulated in NMR.
In the short-time regime, cf. Fig.~\ref{fig:Dynamics}(b),
there are Rabi-like oscillations of coherent energy transfer
between the two levels with the highest energies,
because there is a strong coupling between these two levels.
Furthermore, the oscillation quickly damps as the energy transfer is irreversible
due to the pure-dephasing noise. After an exponential decay process,
the populations on all levels reach thermal equilibrium. Noticeably,
there are small oscillations for the two lowest-energy levels as a result of their strong coupling.
For each point in the quantum simulation, the data is averaged over $M$ random realizations.
For a given realization, the system undergoes a coherent evolution
by applying a time-dependent magnetic field with fluctuating phases, as shown in Eqs.~(\ref{eq:Bz},\ref{eq:phi}). However, for an ensemble of random realizations,
since each realization experiences a different phase at a given time,
the ensemble average manifests itself as a single realization
undergoing a pure-dephasing noise. In this regard,
the deviation of the NMR simulation
from that predicted by the HEOM decreases
as the number of realizations in the ensemble increases,
cf. the Supplementary Information.
This effect would be more remarkable if $M$ were increased further,
as confirmed by our theoretical simulations in the Supplementary Information.
In conclusion, the dephasing Hamiltonian~(\ref{eq:Hsb}) in photosynthesis is effectively mimicked by
a classical time-fluctuating magnetic field (\ref{eq:Bz},\ref{eq:phi}).

Note that the proof-of-concept NMR experiment presented in this work
goes beyond reproducing the HEOM results. To faithfully simulate
the EET dynamics on a large photosynthetic system, using the HEOM would
be computationally unaffordable. In addition, the HEOM \cite{Tanimura06,Ishizaki09}
is known to encounter difficulties when the spectral density is not in a simple
Drude-Lorentz form.
As shown in the Supplementary Information, the computational cost of HEOM scales
exponentially with the system size and the number of exponentials in the bath
correlation function \cite{Tanimura06,Ishizaki09}.
Our NMR simulations do not have such shortcomings, as it scales polynomially
with respect to the system size \cite{Li17,Lu17}.
Moreover, the quantum simulation algorithm also scales more favorably
in terms of the number of exponentials in the bath correlation function.
With our current approach, in principle, a photosynthetic system with $\sim100$ sites (e.g. the PSI complex)
requires a 7-qubit quantum simulator. This means that the coherent EET
dynamics of a full functional biological unit for photosynthesis
(e.g. the PSII supercomplex that contains $\sim300$ sites) can be
faithfully simulated by a 9-qubit NMR quantum computer,
which is clearly attainable by the NMR technique \cite{Lu17}. Thus,
we believe that NMR quantum simulation has the potential to
help clarify the mysteries of light harvesting in natural photosynthesis.
In this regard, other approaches, which do not take advantage of this scaling provided by encoding multiple sites into a single qubit \cite{Potonik17},
currently lack the required size to simulate such large photosynthetic systems, and thus may also benefit from the approach we develop here.


In this paper, the photosynthetic energy transfer is experimentally
simulated in NMR. As a prototype, a two-qubit NMR system is utilized
to demonstrate both the coherent oscillations at the short-time regime
and the steady-state thermalization at the long-time.
By using the GRAPE technique, an arbitrary photosynthetic system can
be faithfully mapped to the NMR system. Besides,
the effect on EET can be effectively
mimicked by a set of well-designed pulses, which act as a classical pure-dephasing noise.
The quantum simulation of photosynthetic energy transfer in NMR would probably facilitate the investigation of quantum coherent effects on the EET
and more clear design principles for artificial
light-harvesting devices together with structured baths~\cite{Ai13,Rey13}.

A recent work \cite{Potonik17} experimentally verified that a structured bath can optimize the energy transfer in EET \cite{Rey13}. However, here we showed we can use NMR and bath-engineering
techniques to efficiently mimic photosynthetic light harvesting in large systems with arbitrary system structure and bath spectral density. Future extensions of our approach include encoding quantum properties of the environment in ancillary qubits.

\bigskip\noindent
\textbf{Acknowledgements}
We thank Xin-Long Zhen, and Ke-Ren Li for discussions.
GLL was supported by National Natural Science Foundation of China under Grant Nos.~11175094 and~91221205, and the National Basic Research Program of China under Grant No.~2015CB921001.
FGD was supported by the National Natural Science Foundation of China under Grant No.~11474026
and No.~11674033.
FN was supported by the MURI Center for Dynamic Magneto-Optics via the AFOSR Award No.~FA9550-14-1-0040, the Japan Society for the Promotion of Science (KAKENHI), the IMPACT program of JST, JSPS-RFBR grant No.~17-52-50023, CREST grant No.~JPMJCR1676, and RIKEN-AIST Challenge Research Fund. QA was supported by the National Natural Science Foundation of China under Grant No.~11505007.

\bigskip\noindent
\textbf{Author Contributions}
All work was carried out under the supervision of G.L.L., F.G.D., Y.C.C., and F.N.
B.X.W. and T.X. performed the experiments.
Y.C.C., B.X.W., Q.A., M.J.T. and N.L. analysed the experimental data.
M.J.T., Q.A. and N.L. wrote the HEOM simulation code
and performed the numerical simulation.
Y.C.C. and Q.A. contributed the theory.
All authors contributed to writing the manuscript.

\bigskip\noindent
\textbf{Author Information}
Reprints and permissions information is available at www.nature.com/reprints.
The authors declare no competing financial interests.
Readers are welcome to comment on the online version of the paper. Correspondence
and requests for materials should be addressed to G.L.L.(gllong@tsinghua.edu.cn)
and F.G.D.(fgdeng@bnu.edu.cn).

\section*{Methods}
\noindent
\textbf{Physical parameters.} The Hamiltonian implemented in NMR simulation $H_{\textrm{NMR}}$ is the same as the photosynthetic Hamiltonian $H_{\textrm{EET}}$ but the unit cm$^{-1}$ divided by $3\times10^4/\pi$. The diagonal terms are
$H^{11}_{\textrm{NMR}}=2\pi\times650$~kHz, $H^{22}_{\textrm{NMR}}=2\pi\times645$~kHz,
$H^{33}_{\textrm{NMR}}=2\pi\times615$~kHz, and $H^{44}_{\textrm{NMR}}=2\pi\times610$~kHz.
The inter-level couplings are the nearest neighboring couplings $H^{12}_{\textrm{NMR}}=H^{34}_{\textrm{NMR}}=2\pi\times6.3040$~kHz,
$H^{23}_{\textrm{NMR}}=2\pi\times6.5950$~kHz, the next-nearest-neighboring couplings $H^{13}_{\textrm{NMR}}=H^{24}_{\textrm{NMR}}=2\pi\times0.8059$~kHz,
and the coupling between the two ends $H^{14}_{\textrm{NMR}}=2\pi\times0.2370$~kHz.
The temperature of the photosynthetic energy transfer and NMR experiment are respectively $T_{\textrm{EET}}=3\times10^4$~K and $T_{\textrm{NMR}}=5\times10^{-5}$~K.
The reorganization energy of the bath is $\lambda_\textrm{EET}=0.2$~cm$^{-1}$.
The cutoff frequency of the bath is $\gamma_\textrm{EET}=900$~cm$^{-1}$.
Both the Hamiltonian and bath parameters for NMR experiment are
those for the photosynthetic energy transfer
which are scaled down by a factor of $3\times10^8/\pi$.

\smallskip\noindent
\textbf{Initialization.} Starting from the thermal equilibrium state,
we prepare a pseudo-pure state~\cite{Gershenfeld97,Cory97}
using the spatial-average technique~\cite{Cory97}.

\smallskip\noindent
\textbf{Measurement.} The goal is to acquire probability
distributions of four states $|00\rangle$, $|01\rangle$,
$|10\rangle$, and $|11\rangle$ after the pulse sequences that we
designed are implemented on the two-qubit NMR system, namely, four
diagonal values of the final density matrix. The density matrices of the
output states are reconstructed completely via quantum state
tomography~\cite{Xin17}. Therefore, the density
matrix of the system can be estimated from ensemble averages of a
set of observables. For the one-qubit system, the observable set is
$\{\sigma_i\}(i=0, 1, 2,3)$. Here, $\sigma_0=I$, $\sigma_1=\sigma_x$,
$\sigma_2=\sigma_y$, $\sigma_3=\sigma_z$. For the two-qubit system, the observable
set is $\{\sigma_i\otimes\sigma_j\}(i, j=0, 1, 2, 3)$. In our
experiments, the complete density matrix tomography is not
necessary. All we need is to perform two experiments in which the
reading-out pulses $\exp(-i\pi \sigma_y/4)\otimes I$ and $I\otimes\exp(-i\pi \sigma_y/4)$ are
respectively implemented on the final states of ${}^{1}$H and
${}^{13}$C and the corresponding qubits that need to be observed are
respectively ${}^{1}$H and ${}^{13}$C. The points in
Figs.~\ref{fig:Dynamics} is obtained by averaging over $M=150$ random realizations.


\newpage

\centerline{{\bf\large Supplementary information for "Quantum simulation of photosynthetic energy transfer"}}
\medskip
\bigskip

\section{Ramsey-like Dynamics in Photosynthesis}
\label{sec:Photosynthesis}

In this section, we provide a detailed derivation for the Ramsey-like dynamics in photosynthesis \cite{Cheng09,Lambert13}. We assume the total Hamiltonian to be
\begin{equation}\label{eq1}
H=\varepsilon_{D}\vert1\rangle\langle1\vert+\varepsilon_{A}\vert2\rangle\langle2\vert+\sum_{k}\omega_{k}a_{k}^{\dagger}a_{k}+\sum_{k}\omega_{k}b_{k}^{\dagger}b_{k}
    +\vert1\rangle\langle1\vert\sum_{k}g_{k}\left(a_{k}^{\dagger}+a_{k}\right)+\vert2\rangle\langle2\vert\sum_{k}g_{k}\left(b_{k}^{\dagger}+b_{k}\right),
\end{equation}
where we have assumed that the dimer is subject to two local harmonic-oscillator baths with the same parameters.

In the experiment, the system is initialized to $\vert1\rangle$ and followed by a $\pi/2$ pulse, i.e.,
\begin{equation}
\vert\psi\left(0\right)\rangle=\exp\left(i\frac{\pi}{4}\sigma_{x}\right)\vert1\rangle=\left(\vert1\rangle+i\vert2\rangle\right)/\sqrt{2}.
\end{equation}
And the bath is in thermal equilibrium, i.e.,
\begin{equation}
\rho_{B}=\prod_{k}^{\otimes}\frac{1}{Z_{k}}\sum_{n=0}^{\infty}\exp(-n\beta\omega_{k})\vert n\rangle_{a_{k}}\langle n\vert\otimes\prod_{k^{\prime}}^{\otimes}\frac{1}{Z_{k^{\prime}}}\sum_{m=0}^{\infty}\exp(-m\beta\omega_{k^{\prime}})\vert m\rangle_{b_{k^{\prime}}}\langle m\vert,
\end{equation}
where the partition function of $k$th bath mode is
\begin{equation}
Z_{k}=\frac{1}{1-e^{-\beta\omega_{k}}}.
\end{equation}
Then, the system evolves under the Hamiltonian~(\ref{eq1}) for a time interval $t$ and thus results in
\begin{eqnarray}
\rho\left(t\right)&=&\textrm{Tr}_{B}\left[U\left(t\right)\vert\psi\left(0\right)\rangle\langle\psi\left(0\right)\vert\otimes\rho_{B}U^{\dagger}\left(t\right)\right]
    \notag \\
    &=&\left(\begin{array}{cc}a\left(t\right) & b\left(t\right) \\ b^{*}\left(t\right) & 1-a\left(t\right)\end{array}\right).
\end{eqnarray}

Finally, after applying a reverse $\pi/2$ pulse, we measure the population of $\vert1\rangle$ in the final state, i.e.,
\begin{equation}
\begin{split}
\rho\left(t_{f}\right)&=\exp\left(-i\pi\sigma_{x}/4\right)\rho\left(t\right)\exp\left(i\pi\sigma_{x}/4\right)                   \\
&=\frac{1}{2}\left(\begin{array}{cc}1+i\left(b-b^{*}\right) & \left(b+b^{*}\right)-i\left(1-2a\right) \\
        \left(b+b^{*}\right)+i\left(1-2a\right) & 1-i\left(b-b^{*}\right)\end{array}\right).
\end{split}
\end{equation}
Thus, the populations of $\vert1\rangle$ reads
\begin{equation}
P_{1}\left(t_{f}\right)=\frac{1}{2}\left[1+i\left(b-b^{*}\right)\right].
\end{equation}
The off-diagonal element can be calculated as
\begin{eqnarray}
b\left(t\right)&=&\textrm{Tr}_{B}\left[U\left(t\right)\vert1\rangle\langle2\vert\otimes\rho_{B}U^{\dagger}\left(t\right)\right] \notag\\
&=&\textrm{Tr}_{B}\left[e^{-iH_{1}t}\rho_{B}e^{iH_{2}t}\right]\notag\\
&=&\exp[-i\left(\varepsilon_{D}-\varepsilon_{A}\right)t]\prod_{k,k^{\prime}}I_{k}^{(a)}I_{k^{\prime}}^{(b)},
\end{eqnarray}
where
\begin{eqnarray}
H_{1}&=&\varepsilon_{D}+\sum_{k}\omega_{k}a_{k}^{\dagger}a_{k}+\sum_{k}\omega_{k}b_{k}^{\dagger}b_{k}+\sum_{k}g_{k}\left(a_{k}^{\dagger}+a_{k}\right),     \\
H_{2}&=&\varepsilon_{A}+\sum_{k}\omega_{k}a_{k}^{\dagger}a_{k}+\sum_{k}\omega_{k}b_{k}^{\dagger}b_{k}+\sum_{k}g_{k}\left(b_{k}^{\dagger}+b_{k}\right),     \\
I_{k}^{(a)}&=&\textrm{Tr}_{B}\left[\exp\left(-i\left[\omega_{k}a_{k}^{\dagger}a_{k}+g_{k}\left(a_{k}^{\dagger}+a_{k}\right)\right]t\right)\frac{1}{Z_{k}}
        \sum_{n=0}^{\infty}\exp\left(-n\beta\omega_{k}\right)\vert n\rangle_{a_{k}}\langle n\vert\exp\left(i\omega_{k}ta_{k}^{\dagger}a_{k}\right)\right],  \\
I_{k}^{(b)}&=&\textrm{Tr}_{B}\left[\exp\left(-i\omega_{k}tb_{k}^{\dagger}b_{k}\right)\frac{1}{Z_{k}}\sum_{n=0}^{\infty}\exp\left(-n\beta\omega_{k}\right)\vert n\rangle_{b_{k}}\langle n\vert
        \exp\left(i\left[\omega_{k}b_{k}^{\dagger}b_{k}+g_{k}\left(b_{k}^{\dagger}+b_{k}\right)\right]t\right)\right]\notag\\
&=&\left[I_{k}^{(a)}\right]^{*}.
\end{eqnarray}

Hereafter, we shall explicitly give the expression of $I_{k}^{(a)}$ as
\begin{eqnarray}
I_{k}^{(a)}&=&\frac{1}{Z_{k}}\textrm{Tr}_{B}\left[\exp\left(-i\left[\omega_{k}a_{k}^{\dagger}a_{k}+g_{k}\left(a_{k}^{\dagger}+a_{k}\right)\right]t\right)
    \sum_{n=0}^{\infty}\exp\left(-\beta\omega_{k}a_{k}^{\dagger}a_{k}\right)\vert n\rangle_{a_{k}}\langle n\vert\exp\left(i\omega_{k}ta_{k}^{\dagger}a_{k}\right)\right]  \notag      \\
&=&\frac{1}{Z_{k}}\textrm{Tr}_{B}\left[\exp\left(-i\left[\omega_{k}a_{k}^{\dagger}a_{k}+g_{k}\left(a_{k}^{\dagger}+a_{k}\right)\right]t\right)
    \exp\left(-\beta\omega_{k}a_{k}^{\dagger}a_{k}\right)\exp\left(i\omega_{k}ta_{k}^{\dagger}a_{k}\right)\right]   \notag \\
&=&\frac{1}{Z_{k}}\textrm{Tr}_{B}\left[D_{k}^{\dagger}\!\left(\frac{g_{k}}{\omega_{k}}\right)\exp\left(-i\omega_{k}ta_{k}^{\dagger}a_{k}\right)\exp\left(i\frac{g_{k}^{2}}{\omega_{k}}t\right)
    D_{k}\!\left(\frac{g_{k}}{\omega_{k}}\right)\exp\left(-\beta\omega_{k}a_{k}^{\dagger}a_{k}\right)\exp\left(i\omega_{k}ta_{k}^{\dagger}a_{k}\right)\right],
\end{eqnarray}
where the displacement operator is
\begin{equation}
D_{k}(\alpha)=\exp\left[\alpha\left(a_{k}^{\dagger}-a_{k}\right)\right].
\end{equation}
By using the identity
\begin{equation}
\exp\left(i\omega_{k}ta_{k}^{\dagger}a_{k}\right)a_{k}\exp\left(-i\omega_{k}ta_{k}^{\dagger}a_{k}\right)=a_{k}\exp\left(-i\omega_{k}t\right),
\end{equation}
$I_{k}^{(a)}$ is simplified as
\begin{eqnarray}
I_{k}^{(a)}&=&\frac{1}{Z_{k}}\exp\left(i\frac{g_{k}^{2}}{\omega_{k}}t\right)\textrm{Tr}_{B}\left[\exp\left(i\omega_{k}ta_{k}^{\dagger}a_{k}\right)D_{k}^{\dagger}\!\left(\frac{g_{k}}{\omega_{k}}\right)
    \exp\left(-i\omega_{k}ta_{k}^{\dagger}a_{k}\right)D_{k}\!\left(\frac{g_{k}}{\omega_{k}}\right)\exp\left(-\beta\omega_{k}a_{k}^{\dagger}a_{k}\right)\right]        \notag \\
&=&\frac{1}{Z_{k}}\exp\left(i\frac{g_{k}^{2}}{\omega_{k}}t\right)\textrm{Tr}_{B}\left[\exp\left[\frac{g_{k}}{\omega_{k}}\left(a_{k}e^{-i\omega_{k}t}-a_{k}^{\dagger}e^{i\omega_{k}t}\right)\right]
    \exp\left[\frac{g_{k}}{\omega_{k}}\left(a_{k}^{\dagger}-a_{k}\right)\right]\exp\left(-\beta\omega_{k}a_{k}^{\dagger}a_{k}\right)\right]                         \notag \\
&=&\frac{1}{Z_{k}}\exp\left(i\frac{g_{k}^{2}}{\omega_{k}}t-\frac{ig_{k}^{2}}{\omega_{k}^{2}}\sin\omega_{k}t\right)
    \textrm{Tr}_{B}\left[\exp\left\{\frac{g_{k}}{\omega_{k}}\left[a_{k}^{\dagger}\left(1-e^{i\omega_{k}t}\right)+a_{k}\left(e^{-i\omega_{k}t}-1\right)\right]\right\}
    \exp\left(-\beta\omega_{k}a_{k}^{\dagger}a_{k}\right)\right],
\end{eqnarray}
where in the last line we have used the Baker-Hausdorff formula \cite{Sakurai11}
\begin{equation}
e^{A}e^{B}=e^{\left[A,B\right]/2}e^{A+B}.
\end{equation}
Then, we apply the identity
\begin{equation}
\textrm{Tr}_{B}\left[\exp\left(r_{1}a_{k}+r_{2}a_{k}^{\dagger}\right)\frac{\exp\left(-\beta\omega_{k}a_{k}^{\dagger}a_{k}\right)}{Z_{k}}\right]
        =\exp\left[\frac{1}{2}r_{1}r_{2}\coth\!\left(\frac{\beta\omega_{k}}{2}\right)\right]
\end{equation}
to the above equation, we obtain $I_{k}^{(a)}$ as
\begin{eqnarray}
I_{k}^{(a)}&=&\exp\left[i\frac{g_{k}^{2}}{\omega_{k}}t-\frac{ig_{k}^{2}}{\omega_{k}^{2}}\sin\omega_{k}t+\frac{g_{k}^{2}}{2\omega_{k}^{2}}\left(1-e^{i\omega_{k}t}\right)
        \left(e^{-i\omega_{k}t}-1\right)\coth\!\left(\frac{\beta\omega_{k}}{2}\right)\right]           \notag \\
&=&\exp\left[i\frac{g_{k}^{2}}{\omega_{k}}t-\frac{ig_{k}^{2}}{\omega_{k}^{2}}\sin\omega_{k}t+\frac{g_{k}^{2}}{\omega_{k}^{2}}\left(\cos\omega_{k}t-1\right)
        \coth\left(\frac{\beta\omega_{k}}{2}\right)\right]    \notag  \\
&=&\exp\left\{-\frac{g_{k}^{2}}{\omega_{k}^{2}}\left[\left(1-\cos\omega_{k}t\right)\coth\left(\frac{\beta\omega_{k}}{2}\right)+i\left(\sin\omega_{k}t-\omega_{k}t\right)\right]\right\}  \notag \\
&=&I_{k}^{(b)*},
\end{eqnarray}
By inserting $I_{k}^{(a)}$ into $b(t)$, we have
\begin{eqnarray}
b(t)&=&\exp[-i\left(\varepsilon_{D}-\varepsilon_{A}\right)t]\prod_{k}\exp\left\{-\frac{g_{k}^{2}}{\omega_{k}^{2}}
        \left[\left(1-\cos\omega_{k}t\right)\coth\!\left(\frac{\beta\omega_{k}}{2}\right)+i\left(\sin\omega_{k}t-\omega_{k}t\right)\right]\right\}             \notag \\
&\quad & \times\prod_{k^{\prime}}\exp\left\{-\frac{g_{k^{\prime}}^{2}}{\omega_{k^{\prime}}^{2}}\left[\left(1-\cos\omega_{k^{\prime}}t\right)
        \coth\left(\frac{\beta\omega_{k^{\prime}}}{2}\right)-i\sin\left(\sin\omega_{k^{\prime}}t-\omega_{k^{\prime}}t\right)\right]\right\}                  \notag \\
&=&\exp\left[-i\left(\varepsilon_{D}-\varepsilon_{A}\right)t-g(t)-g^{*}\!(t)\right]          \notag \\
&=&\exp\left\{-i\left(\varepsilon_{D}-\varepsilon_{A}\right)t-2\textrm{Re}[g(t)]\right\},
\end{eqnarray}
where the lineshape function reads
\begin{equation}
g(t)=\sum_{k}\frac{g_{k}^{2}}{\omega_{k}^{2}}\left[\left(1-\cos\omega_{k}t\right)\coth\!\left(\frac{\beta\omega_{k}}{2}\right)+i\left(\sin\omega_{k}t-\omega_{k}t\right)\right].
\end{equation}
The population of $\vert1\rangle$ reads
\begin{eqnarray}
P_{1}(t_{f})=\frac{1}{2}\left\{1+e^{-2\textrm{Re}[g(t_f)]}\cos(\varepsilon_{D}-\varepsilon_{A})t_f\right\}.
\label{P1}
\end{eqnarray}

Since the spectral density is defined as
\begin{eqnarray}
J\left(\omega\right)&=&\sum_{k}g_{k}^{2}\;\delta(\omega-\omega_{k}) \notag\\
 &=&\int d\omega_{k}\;\rho\left(\omega_{k}\right)g_{k}^{2}\;\delta(\omega-\omega_{k}) \notag\\
 &=&\rho(\omega_{k})g_{k}^{2}\vert_{\omega_{k}=\omega}
\end{eqnarray}
with $\rho(\omega_{k})$ being density of states of bath, the lineshape function can explicitly given as
\begin{eqnarray}\label{eq24}
g\left(t\right)&=&\int d\omega_{k}\;\rho(\omega_{k})\frac{g_{k}^{2}}{\omega_{k}^{2}}\left[\left(1-\cos\omega_{k}t\right)\coth\left(\frac{\beta\omega_{k}}{2}\right)
        +i\sin\left(\sin\omega_{k}t-\omega_{k}t\right)\right] \notag    \\
&=&\int d\omega_{k}\frac{J\left(\omega_{k}\right)}{\omega_{k}^{2}}\left[\left(1-\cos\omega_{k}t\right)\coth\left(\frac{\beta\omega_{k}}{2}\right)
        +i\left(\sin\omega_{k}t-\omega_{k}t\right)\right]  \notag     \\
&=&\int_{0}^{\omega_{c}}d\omega_{k}\frac{2\lambda\Lambda}{\left(\omega_{k}^{2}+\Lambda^{2}\right)\omega_{k}}\left[\left(1-\cos\omega_{k}t\right)\coth\left(\frac{\beta\omega_{k}}{2}\right)
        +i\left(\sin\omega_{k}t-\omega_{k}t\right)\right],
\end{eqnarray}
where we assumed a Debye-form spectral density
\begin{equation}
J\left(\omega\right)=\frac{2\lambda\Lambda\omega}{\omega^{2}+\Lambda^{2}}
\end{equation}
with $\lambda$ and $\Lambda$ being the reorganization energy and cutoff frequency respectively.

By using a Matsubara expansion \cite{Mukamel95}, the lineshape function is explicitly calculated as
\begin{equation}
g(t)=\frac{\lambda}{\Lambda}\left[\cot\left(\frac{\beta\Lambda}{2}-i\right)\right]\left(e^{-\Lambda t}+\Lambda t-1\right)
    +\frac{4\lambda\Lambda}{\beta}\sum_{n=1}^{\infty}\frac{e^{-\nu_{n}t}+\nu_{n}t-1}{\nu_{n}\left(\nu_{n}^{2}-\Lambda^{2}\right)},
\end{equation}
where
\begin{equation}
\nu_{n}=\frac{2\pi n}{\beta}.
\end{equation}

In Sec.~\ref{sec:Classical Noise}, we will demonstrate that in order to simulate the photosynthetic dynamics in NMR, the following relations should be fulfilled
\begin{eqnarray}
\chi(t)=\textrm{Re}[g(t)],      \\
\omega_{L}=\varepsilon_{D}-\varepsilon_{A}.
\end{eqnarray}

\section{Hierarchical Equations of Motion (HEOM)}
\label{sec:HEOM}

The hierarchical equations of motion (HEOM) formalism has become an important method for studying quantum open systems \cite{Tanimura06,Ishizaki091,Ishizaki092}. In this section, we describe the application of the HEOM method for studying the excitation energy transfer (EET) in photosynthetic systems \cite{Ishizaki091,Ishizaki092}.

We discuss the EET dynamics in a photosynthetic complex containing four pigments, and each pigment is modeled by a two-level system. The following Frenkel exciton Hamiltonian \cite{Cheng09,Novoderezhkin10}, studying EET dynamics, consists of three parts,
\begin{equation}
H_{\textrm{tot}}=H_{\textrm{el}}+H_{\textrm{ph}}+H_{\textrm{el-ph}},
\end{equation}
where
\begin{eqnarray}
H_{\textrm{el}}&=&\sum_{j=1}^{4}\varepsilon_{j}\vert j\rangle\langle j\vert+\sum_{j<k}^{4}J_{jk}\left(\vert j\rangle\langle k\vert+\vert k\rangle\langle j\vert\right),    \\
H_{\textrm{ph}}&=&\sum_{j=1}^{4}H_{\textrm{ph},j} \notag\\
&=&\sum_{j=1}^{4}\sum_{m}\omega_{jm}\left(p_{jm}^{2}+q_{jm}^{2}\right)/2,          \\
H_{\textrm{el-ph}}&=&\sum_{j=1}^{4}H_{\textrm{el-ph},j} \notag\\
&=&\sum_{j=1}^{4}V_{j}\mu_{j}.
\end{eqnarray}
In the above, $\vert j\rangle$ represents the state where only the $j$th pigment is in its electronic excited state and all others are in their electronic ground state. Moreover, this
\begin{equation}
\varepsilon_{j}=\varepsilon_{j}^{0}+\lambda_{j}
\end{equation}
is the so-called site energy of the $j$th pigment, where $\varepsilon_{j}^{0}$ is the excited electronic energy of the $j$th pigment in the absence of phonons and $\lambda_{j}$ is the reorganization energy of the $j$th pigment. Furthermore, $J_{jk}$ is the electronic coupling between pigments $i$ and $j$. Also, $\omega_{m}$, $p_{jm}$ and $q_{jm}$ are the frequency, dimensionless coordinate, and conjugate momentum of the $m$th phonon mode, respectively. Here,
\begin{eqnarray}
V_{j}&=&\vert j\rangle\langle j\vert, \\
\mu_{j}&=&-\sum_{m}c_{jm}q_{jm}
\end{eqnarray}
with $c_{jm}$ being the coupling constant between the $j$th pigment and $m$th phonon mode. For simplicity, we assume that the phonon modes associated with different pigments are uncorrelated.

The reduced density operator of the system
\begin{eqnarray}
\rho(t)=\textrm{Tr}_{\textrm{ph}}\left\{\rho_{\textrm{tot}}(t)\right\}
\end{eqnarray}
with $\rho_{\textrm{tot}}$ being the density operator for the total system can adequately describe the EET dynamics. At the initial time $t=0$,
we assume that the total system is in the factorized product state of the form
\begin{eqnarray}
\rho_{\textrm{tot}}(0)=\rho(0)\frac{\exp\left(-\beta H_{\textrm{ph}}\right)}{\textrm{Tr}\exp\left(-\beta H_{\textrm{ph}}\right)}\label{eq:iniCond}
\end{eqnarray}
with
\begin{eqnarray}
\beta=\frac{1}{k_{B}T}.
\end{eqnarray}
In accordance to the vertical Franck-Condon transition \cite{Ishizaki091,Ishizaki092},
the initial condition (\ref{eq:iniCond}) is appropriate in electronic excitation processes. In this work,
we adopt the spectral density of the overdamped Brownian oscillator model,
\begin{eqnarray}
J_{j}(\omega)=\frac{2\lambda_{j}\gamma_{j}\omega}{\omega^{2}+\gamma_{j}^{2}},
\end{eqnarray}
to describe the coupling between the $j$th pigment and the environmental phonons. For this modeling,
the timescale of the phonon relaxation is simply,
\begin{eqnarray}
\tau_{c}=\frac{1}{\gamma_{j}}.
\end{eqnarray}
According to the reorganization dynamics, one can determine the reorganization energy $\lambda_{j}$.

For high temperatures $\beta\gamma_{j}<1$, the following hierarchically coupled equations of motion for the reduced density operator with the overdamped Brownian oscillator model is given by
\begin{equation}\label{eq3}
\frac{\partial}{\partial t}\sigma(\mathbf{n},t)=-\left(i \ell_{e}+\sum_{j=1}^{4}n_{j}\gamma_{j}\right)\sigma(\mathbf{n},t)
    +\sum_{j=1}^{4}\left[\Phi_{j}\sigma(\mathbf{n}_{j+},t)+n_{j}\Theta_{j}\sigma(\mathbf{n}_{j-},t)\right],
\end{equation}
where $\mathbf{n},\mathbf{n}_{j\pm}$ are three sets of nonnegative integers, i.e.,
\begin{eqnarray}
\mathbf{n}&=&\left(n_{1},n_{2},n_{3},n_{4}\right), \\
\mathbf{n}_{j\pm}&=&\left(n_{1},\cdots,n_{j}\pm1,\cdots,n_{4}\right).
\end{eqnarray}
The phonon-induced relaxation operators are written by
\begin{eqnarray}
\Phi_{j}&=&iV_{j}^{\times},        \\
\Theta_{j}&=&i\left(2\lambda_{j}TV_{j}^{\times}-i\lambda_{j}\gamma_{j}V^{o}_{j}\right),
\end{eqnarray}
where
\begin{eqnarray}
O^{\times}f&=&[O,f]=Of-fO, \\
O^{o}f&=&\{O,f\}=Of+fO
\end{eqnarray}
are the hyper-operator notations. In addition,
\begin{equation}\rho(t)=\sigma(\mathbf{0},t),
\end{equation}
and the other $\sigma(\mathbf{n}\neq\mathbf{0},t)$ are auxiliary operators considering the fluctuation and dissipation. The Liouvillian operator $\ell_{e}$ corresponds to the electronic Hamiltonian $H_{e}$.

We terminate Eq.~(\ref{eq3}), when the integers $n_j$'s satisfy
\begin{equation}
N=\sum_{j=1}^{4}n_{j}\gg\frac{\omega_{e}}{\min\left(\gamma_{1},\gamma_{2},\gamma_{3},\gamma_{4}\right)},
\end{equation}
where $\omega_{e}$ is a characteristic frequency of the system dynamics $\ell_{e}$ \cite{Ishizaki091}. The required number of auxiliary density operators $\sigma(\mathbf{n},t)$ is given by
\begin{equation}
\sum_{k=0}^{N}\left(\begin{array}{c}k+4-1 \\ 4-1\end{array}\right)=\frac{\left(4+N\right)!}{4!N!}.
\end{equation}

\section{Dynamics in Classical Pure-dephasing Noise}
\label{sec:Classical Noise}

\subsection{General Case}
In this section, inspired by Ref.~\cite{Green13}, we provide a detailed calculation for the dynamics in the classical pure-dephasing noise.
The total Hamiltonian
\begin{equation}
H\left(t\right)=H_{0}\left(t\right)+H_{c}\left(t\right)
\end{equation}
is divided into two parts, i.e. the control Hamiltonian
\begin{equation}
H_{c}\left(t\right)=\vec{h}\left(t\right)\cdot\vec{\sigma},
\end{equation}
and the noise Hamiltonian
\begin{equation}
H_{0}\left(t\right)=\vec{\beta}\left(t\right)\cdot\vec{\sigma},
\end{equation}
where
\begin{eqnarray}
\vec{h}\left(t\right)&=&\left(h_{x}\left(t\right),h_{y}\left(t\right),h_{z}\left(t\right)\right), \\
\vec{\beta}\left(t\right)&=&\left(\beta_{x}\left(t\right),\beta_{y}\left(t\right),\beta_{z}\left(t\right)\right), \\
\vec{\sigma}&=&\left(\sigma_{x},\sigma_{y},\sigma_{z}\right).
\end{eqnarray}

In the rotating frame with respect to
\begin{equation}
U_{c}\left(t\right)=\mathcal{T}\exp\left[-i\int_{0}^{t}d\tau H_{c}\left(\tau\right)\right],
\end{equation}
the noise Hamiltonian reads
\begin{equation}
\widetilde{H}_{0}\left(t\right)=U_{c}^{\dagger}\left(t\right)H_{0}\left(t\right)U_{c}\left(t\right).
\end{equation}
And the propagator in this frame is correspondingly
\begin{equation}
\widetilde{U}\left(t\right)=\mathcal{T}\exp\left[-i\int_{0}^{t}d\tau\widetilde{H}_{0}\left(\tau\right)\right].
\end{equation}
Therefore, transformed back to the Schr\"{o}dinger picture, the propagator is written as
\begin{equation}
U\left(t\right)=U_{c}\left(t\right)\widetilde{U}\left(t\right).
\end{equation}
Let us now consider
\begin{eqnarray}
\widetilde{H}_{0}\left(t\right)&=&U_{c}^{\dagger}\left(t\right)\vec{\beta}\left(t\right)\cdot\vec{\sigma}U_{c}\left(t\right)     \notag\\
&=&\sum_{i}\beta_{i}\left(t\right)U_{c}^{\dagger}\left(t\right)\sigma_{i}U_{c}\left(t\right)         \notag\\
&=&\sum_{i,j}\beta_{i}\left(t\right)R_{ij}\left(t\right)\sigma_{j},\label{eq:H0}
\end{eqnarray}
where
\begin{equation}
R_{ij}\left(t\right)=\frac{1}{2}\textrm{Tr}\left[U_{c}^{\dagger}\left(t\right)\sigma_{i}U_{c}\left(t\right)\sigma_{j}\right],
\end{equation}
and in the last line of Eq.~(\ref{eq:H0}) we have used the relation
\begin{equation}
\textrm{Tr}\left[\sigma_{i}\sigma_{j}\right]=2\delta_{ij}.
\end{equation}
Hereafter, we shall use the compact definition
\begin{equation}
\overrightarrow{R}=\left(\begin{array}{c} R_{x}\left(t\right) \\ R_{y}\left(t\right) \\ R_{z}\left(t\right)\end{array}\right),
\end{equation}
where
\begin{equation}
R_{i}\left(t\right)=\left(R_{ix}\left(t\right),R_{iy}\left(t\right),R_{iz}\left(t\right)\right).
\end{equation}

When
\begin{eqnarray}
\widetilde{U}\left(t\right)&=&\exp\left[-i\Phi\left(\tau\right)\right], \label{eq:Utilt}\\
\Phi\left(\tau\right)&=&\sum_{\mu=1}^{\infty}\Phi_{\mu}\left(\tau\right),
\end{eqnarray}
according to the Magnus expansion \cite{Blanes09},
we have
\begin{eqnarray}
\Phi_{1}(\tau)&=&\int_{0}^{\tau}dt\widetilde{H}_{0}(t),        \\
\Phi_{2}(\tau)&=&-\frac{i}{2}\int_{0}^{\tau}dt_{1}\int_{0}^{t_{1}}dt_{2}\left[\widetilde{H}_{0}(t_{1}),\widetilde{H}_{0}(t_{2})\right],         \\
\Phi_{3}(\tau)&=&-\frac{1}{6}\int_{0}^{\tau}dt_{1}\int_{0}^{t_{1}}dt_{2}\int_{0}^{t_{2}}dt_{3}\left\{\left[\widetilde{H}_{0}(t_{1}),\left[\widetilde{H}_{0}(t_{2}),
    \widetilde{H}_{0}(t_{3})\right]\right]+\left[\widetilde{H}_{0}(t_{3}),\left[\widetilde{H}_{0}(t_{2}),\widetilde{H}_{0}(t_{1})\right]\right]\right\}.
\end{eqnarray}
By using the identity
\begin{equation}
\left[\overrightarrow{u}\vec{\sigma},\overrightarrow{v}\vec{\sigma}\right]=2i\left(\overrightarrow{u}\times\overrightarrow{v}\right)\vec{\sigma},
\end{equation}
the propagator (\ref{eq:Utilt}) can be rewritten as
\begin{eqnarray}
\widetilde{U}(t)&=&\exp\left[-i\vec{a}(\tau)\cdot\vec{\sigma}\right]\notag\\
&=&\exp\left[-i\sum_{\mu}\vec{a}_{\mu}(\tau)\cdot\vec{\sigma}\right],
\end{eqnarray}
where
\begin{eqnarray}
\vec{a}_{1}(\tau)&=&\sum_{i}\int_{0}^{\tau}dt\;\beta_{i}R_{i}(t),         \\
\vec{a}_{2}(\tau)&=&\sum_{i,j}\int_{0}^{\tau}dt_{1}\int_{0}^{t_{1}}dt_{2}\;\beta_{i}(t_{1})\;\beta_{j}(t_{1})\;\widetilde{R}_{ij}(t_{1},t_{2}),     \\
\vec{a}_{3}(\tau)&=&\frac{2}{3}\sum_{i,j,k}\int_{0}^{\tau}dt_{1}\int_{0}^{t_{1}}dt_{2}\int_{0}^{t_{2}}dt_{3}\;\beta_{i}(t_{1})\;\beta_{j}(t_{2})\;
    \beta_{k}(t_{3})\;\widetilde{R}_{ijk}(t_{1},t_{2},t_{3}),
\end{eqnarray}
with
\begin{eqnarray}
\widetilde{R}_{ij}(t_{1},t_{2})&=&R_{i}(t_{1})\times R_{j}(t_{2}),       \\
\widetilde{R}_{ijk}(t_{1},t_{2},t_{3})&=&R_{i}(t_{1})\times\left[R_{j}(t_{2})\times R_{k}\left(t_{3}\right)\right]
    +R_{k}(t_{3})\times\left[R_{j}(t_{2})\times R_{i}(t_{1})\right].
\end{eqnarray}
To calculate the fidelity of the operation described by
\begin{equation}
U_{c}(t)=\mathcal{T}\exp\left[-i\int_{0}^{t}d\tau H_{c}(\tau)\right],
\end{equation}
we use the Hilbert-Schmidt inner product to measure the fidelity as
\begin{eqnarray}
\mathcal{F}\;(\tau)&=&\frac{1}{4}\left\vert\textrm{Tr}\left[U_{c}^{\dagger}(\tau)U_{c}(\tau)\widetilde{U}\!(\tau)\right]\right\vert^{2}    \notag\\
&=&\frac{1}{4}\left\vert\textrm{Tr}\left[\widetilde{U}\!(\tau)\right]\right\vert^{2}           \notag\\
&=&\frac{1}{4}\left\langle\left\vert\textrm{Tr}\left[\widetilde{U}\!(\tau)\right]\right\vert^{2}\right\rangle         \notag\\
&=&\frac{1}{4}\left\langle\left\vert\textrm{Tr}\exp\left[-i\vec{a}(\tau)\cdot\vec{\sigma}\right]\right\vert^{2}\right\rangle        \notag\\
&=&\frac{1}{4}\left\langle\left\vert\textrm{Tr}\sum_{n=0}^{\infty}\frac{\left(-i\right)^{n}}{n!}\left(a_{x}\sigma_{x}+a_{y}\sigma_{y}+a_{z}\sigma_{z}\right)\right\vert^{2}\right\rangle     \notag\\
&=&\frac{1}{4}\left\langle\left\vert\textrm{Tr}\left[I+\frac{-1}{2!}a^{2}I+\frac{(-1)^{2}}{4!}a^{4}I+\cdots\right]\right\vert^{2}\right\rangle     \notag\\
&=&\langle\cos^{2}a\rangle   \notag   \\
&=&\frac{1}{2}\left[\langle\cos 2a\rangle+1\right]  \notag     \\
&=&\frac{1}{2}\left[1+\sum_{n=0}^{\infty}\left(-1\right)^{m}\frac{2^{2m}}{\left(2m\right)!}\langle a^{2m}\rangle\right],
\end{eqnarray}
where $a$ is the modulus of the vector $\vec{a}(\tau)$, and $\langle\cdots\rangle$ is averaged over all possible noise trajectories.
In the above equation, the lowest order term is
\begin{eqnarray}
\langle a^{2}\rangle&=&\langle aa^{T}\rangle  \notag   \\
&=&\sum_{\mu\nu}\langle a_{\mu}a_{\nu}^{T}\rangle  \notag  \\
&=&\langle a_{1}a_{1}^{T}\rangle+\langle a_{2}a_{2}^{T}\rangle+\cdots+2\left(\langle a_{1}a_{2}^{T}\rangle+\langle a_{1}a_{3}^{T}\rangle+\langle a_{1}a_{4}^{T}\rangle+\cdots\right).
\end{eqnarray}

Thus, the fidelity can be expanded as
\begin{eqnarray}
\mathcal{F}(\tau)&=&1-\langle a_{1}a_{1}^{T}\rangle-2\langle a_{1}a_{2}^{T}\rangle+\left(-\langle a_{2}a_{2}^{T}\rangle-2\langle a_{1}a_{3}^{T}\rangle
    +\frac{1}{3}\langle a_{1}a_{1}^{T}a_{1}a_{1}^{T}\rangle\right)+\cdots \notag  \\
&\simeq& 1-\langle a_{1}a_{1}^{T}\rangle  \notag   \\
&=&1-\sum_{i,j}\int_{0}^{\tau}dt_{1}\int_{0}^{\tau}dt_{2}\;\langle\beta_{i}(t_{1})\beta_{j}(t_{2})\rangle R_{i}(t_{1})R_{j}^{T}(t_{2}) \notag    \\
&=&1-\sum_{i,j,k}\int_{0}^{\tau}dt_{1}\int_{0}^{\tau}dt_{2}\;\langle\beta_{i}(t_{1})\beta_{j}(t_{2})\rangle R_{ik}(t_{1})R_{kj}^{T}(t_{2}) \notag    \\
&=&1-\sum_{i,j,k}\int_{0}^{\tau}dt_{1}\int_{0}^{\tau}dt_{2}\;\langle\beta_{i}(t_{1})\beta_{j}(t_{2})\rangle R_{ik}(t_{1})R_{jk}^{*}(t_{2}),
\end{eqnarray}
where we have used the relation
\begin{equation}
R_{kj}^{T}(\tau)=R_{jk}^{*}(\tau).
\end{equation}

By introducing the Fourier transform of the cross-power spectrum
\begin{equation}
\langle\beta_{i}(t_{1})\beta_{j}(t_{2})\rangle=\frac{1}{2\pi}\int_{-\infty}^{\infty}d\omega\; S_{ij}(\omega)e^{i\omega(t_{2}-t_{1})},
\end{equation}
we have
\begin{eqnarray}
\mathcal{F}(\tau)&=&1-\frac{1}{2\pi}\sum_{i,j,k}\int_{0}^{\tau}dt_{1}\int_{0}^{\tau}dt_{2}\int_{-\infty}^{\infty}d\omega\; S_{ij}(\omega)e^{i\omega(t_{2}-t_{1})}
    R_{ik}(t_{1})R_{jk}^{*}(t_{2})  \notag      \\
&=&1-\frac{1}{2\pi}\sum_{i,j,k}\int_{-\infty}^{\infty}\frac{d\omega}{\omega^{2}}S_{ij}(\omega)\left(-i\omega\right)\int_{0}^{\tau}dt_{1}\;R_{ik}(t_{1})e^{-i\omega t_{1}}
    \left(i\omega\right)\int_{0}^{\tau}dt_{2}\;R_{jk}^{*}(t_{2})e^{i\omega t_{2}}   \notag   \\
&=&1-\frac{1}{2\pi}\sum_{i,j,k}\frac{d\omega}{\omega^{2}}S_{ij}(\omega)R_{ik}(\omega)R_{jk}^{*}(\omega),
\end{eqnarray}
where we have defined
\begin{eqnarray}
R_{ik}(\omega)&=&-i\omega\int_{0}^{\tau}dt\; R_{ik}(t)e^{-i\omega t},         \\
R_{ik}^{*}(\omega)&=&i\omega\int_{0}^{\tau}dt\; R_{ik}^{*}e^{i\omega t}.
\end{eqnarray}

\subsection{Ramsey Fringes}
\label{sec:Ramsey}

In the following, we shall consider a special case where $[H_0(t),H_c(t)]=0$. At the end of this subsection, we will provide the deviation for Ramsey-interferometer experiment. In this case, we assume the total Hamiltonian as
\begin{equation}\label{eq2}
H(t)=\frac{\omega_{L}}{2}\sigma_{z}+\beta_{z}(t)\sigma_{z}.
\end{equation}
Thus, the control Hamiltonian is
\begin{equation}
H_{c}=\frac{\omega_{L}}{2}\sigma_{z},
\end{equation}
and the noise Hamiltonian is
\begin{equation}
H_{0}(t)=\beta_{z}(t)\sigma_{z}.
\end{equation}

In the rotating frame with respect to
\begin{equation}
U_{c}(t)=\exp\left(-i\frac{\omega_{L}t}{2}\sigma_{z}\right),
\end{equation}
the noise Hamiltonian reads
\begin{eqnarray}
\widetilde{H}_{0}(t)&=&U_{c}^{\dagger}(t)H_{0}(t)U_{c}(t)\notag\\
&=&H_{0}(t)\notag\\
&=&\beta_{z}(t)\sigma_{z},
\end{eqnarray}
because
\begin{equation}
\left[H_{c},H_{0}(t)\right]=0.
\end{equation}

And the propagator in this frame is correspondingly
\begin{equation}
\widetilde{U}=\exp\left[-i\int_{0}^{t}d\tau\; \beta_{z}(\tau)\sigma_{z}\right].
\end{equation}
Therefore, transformed back to the Schr\"{o}dinger picture, the propagator is written as
\begin{eqnarray}
U(t)&=&U_{c}(t)\widetilde{U}(t)\notag\\
&=&\exp\left(-i\frac{\omega_{L}t}{2}\sigma_{z}\right)\exp\left[-i\int_{0}^{t}d\tau\; \beta_{z}(\tau)\sigma_{z}\right].
\end{eqnarray}
In the experiment, the system is initialized to $\vert0\rangle$ and followed by a $\pi/2$ pulse, i.e.
\begin{eqnarray}
\vert\psi(0)\rangle&=&\exp\left(i\frac{\pi}{4}\sigma_{x}\right)\vert0\rangle\notag\\
&=&\left(\vert0\rangle+i\vert1\rangle\right)/\sqrt{2}.
\end{eqnarray}
Then, the system evolves under the Hamiltonian~(\ref{eq2}) for a time interval $t$ and thus results in
\begin{eqnarray}
\vert\psi(t)\rangle&=&\frac{1}{\sqrt{2}} U(t)\left(\vert0\rangle+i\vert1\rangle\right)      \notag\\
&=&\frac{1}{\sqrt{2}}\exp\left(-i\frac{\omega_{L}t}{2}\sigma_{z}\right)\exp\left[-i\int_{0}^{t}d\tau\; \beta_{z}(\tau)\sigma_{z}\right]\left(\vert0\rangle+i\vert1\rangle\right)      \notag\\
&=&\frac{1}{\sqrt{2}}\left(e^{-i\phi(t)}\vert0\rangle+ie^{i\phi(t)}\vert1\rangle\right),
\end{eqnarray}
where
\begin{eqnarray}
\phi(t)&=&\phi_{A}(t)+\phi_{B}(t),     \\
\phi_{A}(t)&=&\frac{\omega_{L}t}{2},                         \\
\phi_{B}(t)&=&\int_{0}^{t}d\tau\;\beta_{z}(\tau).
\end{eqnarray}
Finally, after applying a reverse $\pi/2$ pulse, we measure the population of $\vert0\rangle$ in the final state, i.e.
\begin{eqnarray}
\vert\psi_{f}(t)\rangle&=&\frac{1}{\sqrt{2}}e^{-i\pi\sigma_{z}/4}\left(e^{-i\phi(t)}\vert0\rangle+ie^{i\phi(t)}\vert1\rangle\right)       \notag\\
&=&\cos\phi(t)\vert0\rangle+\sin\phi(t)\vert1\rangle.
\end{eqnarray}
Thus, the population of $\vert0\rangle$ reads
\begin{eqnarray}
P_{0}(t)&=&\cos^{2}\phi(t)   \notag      \\
&=&\frac{1}{2}\left[1+\cos2\phi(t)\right]  \notag   \\
&=&\frac{1}{2}\left[1+\cos2\phi_{A}\langle\cos2\phi_{B}(t)\rangle-\sin2\phi_{A}(t)\langle\sin2\phi_{B}(t)\rangle\right],
\end{eqnarray}
where $\langle\cdots\rangle$ is averaged over all possible random realizations. If we further assume a Gaussian noise,
\begin{equation}
\langle\phi_{B}^{2n-1}(t)\rangle=0
\end{equation}
for any positive integer $n$, $P_{0}(t)$ can be simplified as
\begin{eqnarray}
P_{0}(t)&=&\frac{1}{2}\left[1+\cos2\phi_{A}(t)\langle\cos2\phi_{B}(t)\rangle\right]         \notag\\
&=&\frac{1}{2}\left[1+\cos2\phi_{A}(t)\sum_{n=0}^{\infty}\left(-1\right)^{n}\frac{2^{2n}}{(2n)!}\langle\phi_{B}^{2n}(t)\rangle\right].
\end{eqnarray}

For the lowest nontrivial order $n=1$, we have
\begin{eqnarray}
\chi(t)&=&\langle\phi_{B}^{2}(t)\rangle   \notag     \\
&=&\int_{0}^{t}d\tau_{1}\int_{0}^{t}d\tau_{2}\;\langle\beta_{z}(\tau_{1})\beta_{z}(\tau_{2})\rangle       \notag\\
&=&\frac{1}{2\pi}\int_{0}^{t}d\tau_{1}\int_{0}^{t}d\tau_{2}\;\int_{-\infty}^{\infty}d\omega\; S_{zz}\left(\omega\right)e^{i\omega\left(\tau_{2}-\tau_{1}\right)} \notag    \\
&=&\frac{1}{2\pi}\int_{-\infty}^{\infty}d\omega \; S_{zz}\left(\omega\right)\int_{0}^{t}d\tau_{1}\;e^{-i\omega\tau_{1}}\int_{0}^{t}d\tau_{2} \; e^{i\omega\tau_{2}}  \notag    \\
&=&\frac{4}{2\pi}\int_{-\infty}^{\infty}\frac{d\omega}{\omega^{2}}S_{zz}(\omega)\sin^{2}\frac{\omega t}{2},
\end{eqnarray}
where we have introduced the Fourier transform of $\langle\beta_{z}(\tau_{1})\beta_{z}(\tau_{2})\rangle$ as
\begin{equation}
S_{zz}(\omega)=\int_{-\infty}^{\infty}dt\;\langle\beta_{z}(0)\beta_{z}(t)\rangle e^{i\omega t}
\end{equation}
with $\langle\beta_{z}(\tau_{1})\beta_{z}(\tau_{2})\rangle$ being only dependent on the time interval $\tau_{2}-\tau_{1}$.

For the order with $n=2$, we have
\begin{eqnarray}
\langle\phi_{B}^{4}(t)\rangle&=&\int_{0}^{t}d\tau_{1}\int_{0}^{t}d\tau_{2}\int_{0}^{t}d\tau_{3}\int_{0}^{t}d\tau_{4}\;
        \langle\beta_{z}(\tau_{1})\beta_{z}(\tau_{2})\beta_{z}(\tau_{3})\beta_{z}(\tau_{4})\rangle          \notag    \\
&=&\int_{0}^{t}d\tau_{1}\int_{0}^{t}d\tau_{2}\int_{0}^{t}d\tau_{3}\int_{0}^{t}d\tau_{4}\;\left[\left\langle\beta_{z}(\tau_{1})\beta_{z}(\tau_{2})\right\rangle
        \left\langle\beta_{z}(\tau_{3})\beta_{z}(\tau_{4})\right\rangle+
        \left\langle\beta_{z}(\tau_{1})\beta_{z}(\tau_{3})\right\rangle
        \left\langle\beta_{z}(\tau_{2})\beta_{z}(\tau_{4})\right\rangle    \right.    \notag    \\
&\quad&  +\left.      \left\langle\beta_{z}(\tau_{1})\beta_{z}(\tau_{4})\;\right\rangle
        \left\langle\beta_{z}(\tau_{2})\beta_{z}(\tau_{3})\right\rangle\right]        \notag     \\
&=&3\int_{0}^{t}d\tau_{1}\int_{0}^{t}d\tau_{2}\int_{0}^{t}d\tau_{3}\int_{0}^{t}d\tau_{4}\left\langle\beta_{z}(\tau_{1})\beta_{z}(\tau_{2})\right\rangle
        \left\langle\beta_{z}(\tau_{3})\beta_{z}(\tau_{4})\right\rangle           \notag     \\
&=&3\left[\frac{1}{2\pi}\int_{0}^{t}d\tau_{1}\int_{0}^{t}d\tau_{2}\int_{-\infty}^{\infty}d\omega \; S_{zz}(\omega)e^{i\omega\left(\tau_{2}-\tau_{1}\right)}\right]^{2}   \notag    \\
&=&3\left[\frac{4}{2\pi}\int_{-\infty}^{\infty}\frac{d\omega}{\omega^{2}}S_{zz}(\omega)\sin^{2}\frac{\omega t}{2}\right]^{2}   \notag       \\
&=&3\chi^{2}(t).
\end{eqnarray}

For the order with arbitrary integer $n$, we have
\begin{eqnarray}
\langle\phi_{B}^{2n}(t)\rangle&=&\int_{0}^{t}d\tau_{1}\int_{0}^{t}d\tau_{2}\cdots\int_{0}^{t}d\tau_{2n}
    \left\langle\beta_{z}(\tau_{1})\beta_{z}(\tau_{2})\cdots\beta_{z}(\tau_{2n})\right\rangle          \notag \\
&=&\int_{0}^{t}d\tau_{1}\int_{0}^{t}d\tau_{2}\int_{0}^{t}d\tau_{3}\int_{0}^{t}d\tau_{4}\left[\left\langle\beta_{z}(\tau_{1})\beta_{z}(\tau_{2})\right\rangle
    \left\langle\beta_{z}(\tau_{3})\beta_{z}(\tau_{4})\right\rangle\cdots\left\langle\beta_{z}(\tau_{2n-1})\beta_{z}(\tau_{2n})\right\rangle+
    \cdots\right]    \notag    \\
&=&\frac{\left(2n\right)!}{2^{n}n!}\left[\frac{4}{2\pi}\int_{-\infty}^{\infty}\frac{d\omega}{\omega^{2}}S_{zz}\left(\omega\right)\sin^{2}\frac{\omega t}{2}\right]^{n}  \notag  \\
&=&\frac{\left(2n\right)!}{2^{n}n!}\chi^{n}(t),
\end{eqnarray}
where there are $(2n)!/(2^{n}n!)$ terms in the second line according to Isserlis' theorem if it is a Gaussian noise \cite{Goodman15}. To conclude,
\begin{eqnarray}
P_{0}(t)&=&\frac{1}{2}\left[1+\cos2\Phi_{A}(t)\sum_{n=0}^{\infty}\left(-1\right)^{n}\frac{2^{2n}}{\left(2n\right)!}\frac{\left(2n\right)!}{2^{n}n!}
        \chi^{n}(t)\right]  \notag     \\
&=&\frac{1}{2}\left[1+\cos2\phi_{A}(t)\sum_{n=0}^{\infty}\frac{\left(-2\right)^{n}}{n!}\chi^{n}(t)\right]      \notag\\
&=&\frac{1}{2}\left[1+\cos2\phi_{A}(t)e^{-2\chi(t)}\right].
\label{P0}
\end{eqnarray}
This predicts that before decaying to the steady value $1/2$ in the long run, $P_{0}(t)$ will experience oscillations with frequency $\omega_{L}$.

We assume that
\begin{eqnarray}
\beta_{z}&=&\alpha\sum_{j=1}^{J}F(\omega_{j})\omega_{j}\cos\left(\omega_{j}t+\psi_{j}\right)       \notag\\
&=&\frac{\alpha}{2}\sum_{j=1}^{J}F(\omega_{j})\omega_{j}\left[e^{i\left(\omega_{j}t+\psi_{j}\right)}+e^{-i\left(\omega_{j}t+\psi_{j}\right)}\right],
\end{eqnarray}
where the $\psi_{j}$'s are random numbers. According to Ref.~\cite{Goodman15}, the ensemble average is equivalent to the time average for a wide-sense-stationary random process. In this case, the two-time correlation function reads
\begin{eqnarray}
\left\langle\beta_{z}(t+\tau)\beta_{z}(t)\right\rangle&=&\lim_{T\rightarrow\infty}\frac{1}{2T}\int_{-T}^{T}dt\beta_{z}(t+\tau)\beta_{z}(t) \notag\\
&=&\left(\frac{\alpha}{2}\right)^{2}\lim_{T\rightarrow\infty}\frac{1}{2T}\int_{-T}^{T}dt\;\sum_{j,j^{\prime}}\omega_{j}\omega_{j^{\prime}}F(\omega_{j})
    F(\omega_{j^{\prime}})\left[e^{i\omega_{j}(t+\tau)+i\psi_{j}}+e^{-i\omega_{j}(t+\tau)-i\psi_{j}}\right] \notag\\
    &&\times
    \left[e^{i\omega_{j^{\prime}}t+i\psi_{j^{\prime}}}+e^{-i\omega_{j^{\prime}}t-i\psi_{j^{\prime}}}\right]         \notag\\
&=&\left(\frac{\alpha}{2}\right)^{2}\lim_{T\rightarrow\infty}\frac{1}{2T}\int_{-T}^{T}dt\sum_{j}\left[\omega_{j}F(\omega_{j})\right]^{2}
    \left[e^{i\omega_{j}\tau}+e^{-i\omega_{j}\tau}\right]   \notag    \\
&=&\left(\frac{\alpha}{2}\right)^{2}\sum_{j=1}^J\left[\omega_{j}F(\omega_{j})\right]^{2}\left(e^{i\omega_{j}\tau}+e^{-i\omega_{j}\tau}\right),
\end{eqnarray}
which does not depend on $t$ but $\tau$.

The power spectral density is the Fourier transform of the correlation function, i.e.
\begin{eqnarray}
S_{zz}\left(\omega\right)&=&\int_{-\infty}^{\infty}d\tau \; e^{-i\omega\tau}\langle\beta_{z}\left(t+\tau\right)\beta_{z}\left(t\right)\rangle   \notag \\
&=&\int_{-\infty}^{\infty}d\tau \; e^{-\omega\tau}\left(\frac{\alpha}{2}\right)^{2}\sum_{j}\left[\omega_{j}F\left(\omega_{j}\right)\right]^{2}
        \left(e^{i\omega_{j}\tau}+e^{-i\omega_{j}\tau}\right) \notag \\
&=&\left(\frac{\alpha}{2}\right)^{2}\sum_{j=1}^J\left[\omega_{j}F\left(\omega_{j}\right)\right]^{2}\left[\delta\left(\omega-\omega_{j}\right)+\delta\left(\omega+\omega_{j}\right)\right],
\end{eqnarray}
where
\begin{equation}
\delta(\omega\pm\omega_{j})=\int^{\infty}_{\infty} dt\; e^{i(\omega\pm\omega_{j})t}.
\end{equation}
The power spectral density is a set of equally-spaced peaks with distance $\omega_{0}$ and height $\left(\frac{\alpha}{2}\right)^{2}\left[\omega_{j}F(\omega_{j})\right]^{2}$.

If we set
\begin{equation}
F(\omega_{j})=\frac{1}{\sqrt{\omega_{j}}},
\end{equation}
we have
\begin{equation}
S_{zz}(\omega)=\frac{\alpha^{2}}{4}\sum_{j=1}^J\omega_{j}\left[\delta(\omega-\omega_{j})+\delta(\omega+\omega_{j})\right]
\end{equation}
and thus $S_{zz}(\omega)$ is an Ohmic spectral density of step-function form with cutoff frequency
\begin{eqnarray}
\omega_{J}=J\omega_{0}.
\end{eqnarray}

If we set
\begin{equation}
F(\omega_{j})=\frac{1}{\sqrt{\omega_{j}\left(\omega_{j}^{2}+\gamma^{2}\right)}},
\end{equation}
we have
\begin{equation}
S_{zz}(\omega)=\frac{\alpha^{2}}{4}\sum_{j=1}^J\frac{\omega_{j}}{\omega_{j}^{2}+\gamma^{2}}\left[\delta(\omega-\omega_{j})+\delta(\omega+\omega_{j})\right]
\end{equation}
and thus $S_{zz}(\omega)$ is the Debye-Drude spectral density of the step-function form with cutoff frequency $\omega_{J}$.

The transverse relaxation time $T_{2}$ is defined by
\begin{equation}
2\chi(T_{2})=1.
\end{equation}

For photosynthesis, the decoherence is determined by the real part of the lineshape function
\begin{equation}
g(t)=\frac{\lambda}{\Lambda}\left[\cot\left(\frac{\beta\Lambda}{2}\right)-i\right]\left(e^{-\Lambda t+\Lambda t-1}\right)
    +\frac{4\lambda\Lambda}{\beta}\sum_{n=1}^{\infty}\frac{e^{-\nu_{n}t+\nu_{n}t-1}}{\nu_{n}\left(\nu_{n}^{2}-\Lambda^{2}\right)},
\end{equation}
with spectral density
\begin{equation}
J(\omega)=\frac{2\lambda\Lambda\omega}{\omega^{2}+\Lambda^{2}},
\end{equation}
where
\begin{equation}
\nu_{n}=\frac{2\pi n}{\beta}.
\end{equation}

Therefore, in order to simulate photosynthetic dynamics in NMR, we should relate the following two quantities
\begin{eqnarray}
\label{chi}
\chi(t)&=&\textrm{Re}[g(t)]              
\end{eqnarray}
and cutoff frequencies in two spectra are equal, i.e.
\begin{equation}
\gamma=\Lambda.
\end{equation}

\section{Technique of Artificially Injecting Noise}
Here we introduce a method of artificially injecting noises in NMR and ion trap systems, including dephasing noise and amplitude noise \cite{Soare14,Zhen16}.

\subsection{Dephasing Noise}
Dephasing noise comes from the inhomogeneous and non-static magnetic field in NMR systems. The corresponding Hamiltonian can be written as $\beta_z(t)\sigma_z$
\begin{equation}
\beta_z(t)=\sum_{j=1}^N\alpha_zF(\omega_j)\omega_j\cos(\omega_j t+\phi_j),
\end{equation}
where $\alpha_i(i=x,y,z)$ is the noise amplitude and $\phi_j$ is a random phase. $N\omega_0$ determines the high frequency cutoff and $\omega_0$ is the base frequency with $\omega_j=j\omega_0$. The types of noise rely on the function $F(\omega_j)$.
The two-time correlation function for $\beta_z(t)$ is then written as
\begin{eqnarray}
\langle\beta_z(t+\tau)\beta_z(t)\rangle&=&\lim_{T\rightarrow\infty}\frac{1}{2T}\int_{-T}^T dt\;\beta_z(t+\tau)\beta_z(t) \nonumber \\
&=&(\frac{\alpha_z}{2})^2\lim_{T\rightarrow\infty}\frac{1}{2T}\int_{-T}^T dt\sum_{j,j'}\omega_j\omega_{j'}F(\omega_j)F(\omega_{j'})[e^{i\omega_j(t+\tau)+i\phi_j}+e^{-i\omega_j(t+\tau)-i\phi_j}]\nonumber \\
&&\times[e^{i\omega_{j'}(t)+i\phi_{j'}}+e^{-i\omega_{j'}(t)-i\phi_{j'}}]\nonumber\\
&=&(\frac{\alpha_z}{2})^2\lim_{T\rightarrow\infty}\frac{1}{2T}\int_{-T}^T dt\sum_j[\omega_jF(\omega_j)]^2[e^{i\omega_j\tau}+e^{-i\omega_j\tau}]\nonumber\\
&=&(\frac{\alpha_z}{2})^2\sum_j[\omega_j F(\omega_j)]^2(e^{i\omega_j\tau}+e^{-i\omega_j\tau}),
\end{eqnarray}
which does not depend on $t$ but on $\tau$.
Applying the Wiener-Khintchine theorem \cite{Miller12}, we then obtain the power spectral density which can describe the energy distribution of the stochastic signal in the frequency domain by Fourier transform
\begin{align}
S_z(\omega)&=\int_{-\infty}^{\infty} d\tau\; e^{-i\omega\tau}\langle\beta_z(t+\tau)\beta_z(t)\rangle\nonumber\\
&=\int_{-\infty}^{\infty} d\tau\; e^{-i\omega\tau}(\frac{\alpha_z}{2})^2\sum_j[\omega_j F(\omega_j)]^2(e^{i\omega_j\tau}+e^{-i\omega_j\tau})\nonumber\\
&=\frac{\pi\alpha_z^2}{2}\sum_{j=1}^N[F(\omega_j)\omega_j]^2[\delta(\omega-\omega_j)+\delta(\omega+\omega_j)].
\end{align}
Hence, we can use the model of power spectral density to reverse the noise distribution in the time domain. For instance, if we want to simulate the power spectral density for $S(\omega)\sim\omega^p$, then the modulation function $F(\omega_j) = (\omega_j)^{p/2-1}$.
Taking Eq.~(\ref{chi}) into above,
\begin{equation}\label{eq128}
\chi(t)=\alpha_z^2\sum_{j=1}^N[F(\omega_j)]^2\sin^2\frac{\omega_j t}{2}
\end{equation}

The initial state $\vert\psi(0)\rangle=\alpha|0\rangle+\beta|1\rangle$ with the dephasing noise of the Hamiltonian $\beta_z(t)\sigma_z$ in the rotating frame after time $\tau$ will become
\begin{align}
|\psi(\tau)\rangle&=\text{exp}\left[-i\int_{\tau1}^{\tau2}dt\beta_z(t)\sigma_z/2\right](\alpha|0\rangle+\beta|1\rangle)\nonumber\\
                  &=\text{exp}\left[-i\triangle\theta_{\tau}\frac{\sigma_z}{2}\right]\vert\psi(0)\rangle
\end{align}
where $\triangle\theta_{\tau}$ is the integral of $\beta_z(t)$ . Hence we just rotate the angle $\triangle\theta_{\tau}$ along the $z$-axis at the desired point to realize the evolution of the quantum system in the dephasing environment.

\subsection{Amplitude Noise}
Similarly, we can obtain the $\beta_x(t)$ of the amplitude noise as a result of the amplitude
fluctuation of the control field,
\begin{eqnarray}
\beta_x(t)&=&\sum_{j=1}^N\alpha_x F(\omega_j)\sin(\omega_j t+\phi_j),\\
F(\omega_j)&=&(\omega_j)^{p/2},
\end{eqnarray}
and its power spectral density is
\begin{equation}
S(\omega)=\frac{\pi\alpha_x^2}{2}\sum_{j=1}^N[F(\omega_j)]^2[\delta(\omega-\omega_j)+\delta(\omega+\omega_j)]
\end{equation}

\subsection{Parameters for Dephasing Noise of Debye-Drude Form}
For the Debye spectrum of the dephasing noise, taking Re$[g(t)]$ equal to $\chi(t)$, we obtain
\begin{align}
\beta(t)&=\sqrt{\frac{2}{\pi}}\sum_{j=1}^NF(\omega_j)\omega_j\cos(\omega_j t+\phi_j),\\
F(\omega_j)&=\sqrt{\frac{2\lambda\gamma\omega_0 \coth(\frac{\beta\omega_j}{2})}{\omega_j(\omega_j^2+\gamma^2)}}.
\end{align}
In short, as long as we know the power spectral density of the noise, we can then obtain the time-varying $\beta(t)$ and $\chi(t)$. Table~\ref{tab} shows $F(\omega_j)$ for distinct types of dephasing and amplitude noises.

\begin{table}[!ht]\centering
\caption{$F(\omega_j)$ for distinct dephasing and amplitude noises \label{tab}}
\renewcommand{\arraystretch}{4}
\renewcommand{\tabcolsep}{12pt}
\begin{tabular}{|l|c|c|c|c|c|c|c|c|c|}\hline
                            &   \multicolumn{5}{c|}{Dephasing}          & \multicolumn{4}{c|}{Amplitude}     \\  \cline{2-10}
                            &$1/f^{2}$          &$1/f$                  &White              &Ohmic      &Debye      &$1/f^{2}$      &$1/f$      &White      &Ohmic  \\  \hline
$F\left(\omega_{j}\right)$  &$\omega_{j}^{-2}$  &$\omega_{j}^{-3/2}$    &$\omega_{j}^{-1}$  &$\omega_{j}^{-1/2}$
&$\sqrt{\frac{2\lambda\gamma\omega_{0}\coth\left(\beta\omega_{j}/2\right)}{\omega_{j}\left(\omega_{j}^{2}+\gamma^{2}\right)}}$ &$\omega_{j}^{-1}$  &$\omega_{j}^{-1/2}$       &$\omega_{j}^{0}$&$\omega_{j}^{1/2}$        \\  \hline
\end{tabular}
\end{table}

\subsection{Parameters for Dephasing Noise of Arbitrary Form}

For the general spectrum $J(\omega)$ of the dephasing noise, we make Re$[g(t)]$ and $\chi(t)$ be equal according to Eqs.~(\ref{eq24}),~(\ref{chi}),~(\ref{eq128})
\begin{equation}
\alpha_{z}^{2}\sum_{j=1}^{N}\left[F(\omega_{j})\right]^{2}\sin^{2}\frac{\omega_{j}t}{2}=\sum_{j=1}^{N}\frac{J(\omega_{j})\omega_{0}}{\omega_{j}^{2}}
    \left(1-\cos\omega_{j}t\right)\coth\left(\frac{\beta\omega_{j}}{2}\right).
\end{equation}
After simplifying the above formula, we can obtain
\begin{equation}
F(\omega_{j})=\frac{1}{\alpha_{z}}\sqrt{\frac{2J(\omega_{j})\omega_{0}}{\omega_{j}^{2}}\coth\left(\frac{\beta\omega_{j}}{2}\right)}.
\label{eq:F}
\end{equation}

For the B777-complexes in Ref.~\cite{Adolphs06}, of which the spectral density is
\begin{equation}
J(\omega)=\frac{S_0}{s_1+s_2}\sum_{i=1,2}\frac{s_i}{7!2\Omega_i^4}\omega^3e^{-(\omega/\Omega_i)^{1/2}}
\end{equation}
where $s_1=0.8$, $s_2=0.5$, $\Omega_1=0.069$ meV, $\Omega_2=0.24$ meV, $S_0=0.5$,
the corresponding $F(\omega_{j})$ in NMR experiments can be obtained via the above formula (\ref{eq:F}).

\section{Experimental Details and Results}
\begin{figure}[!h]
\begin{center}
\includegraphics[width=14 cm,angle=0]{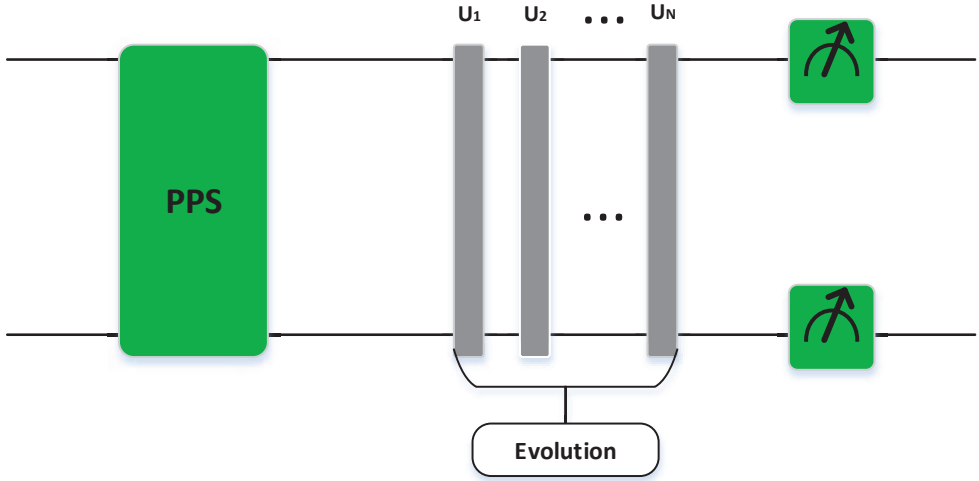}
\caption{ Sequence of the NMR experimental process. It includes three
steps: preparation of the pseudo-pure state, Evolution of the Hamiltonian with Debye noise and measuring the probability distribution of four states.}\label{fig:circuit}
\end{center}
\end{figure}

Experiments are carried out at room temperature using a Bruker Avance III 400 MHz spectrometer. The sample is chloroform dissolved in d6-acetone as a two-qubit NMR quantum processor where H is the first qubit and C is second qubit. The internal Hamiltonian of the two-qubit system can be described as
\begin{equation}
H_{\textrm{int}}=\pi\omega_1\sigma^z_1+\pi\omega_2\sigma^z_2+\frac{\pi}{2}J\sigma^z_1\sigma^z_2,
\end{equation}
where
\begin{eqnarray}
\omega_{1}&=&3206.5~\textrm{Hz}, \\
\omega_{2}&=&7787.9~\textrm{Hz}
\end{eqnarray}
are the chemical shifts of the two spins and
\begin{eqnarray}
J=215.1~\textrm{Hz}
\end{eqnarray}
is the $J$-coupling strength between two spins. The experimental process is divided into three steps, as shown in Fig.~\ref{fig:circuit}.

\begin{figure}[!h]
\begin{center}
\includegraphics[width=8 cm,angle=0]{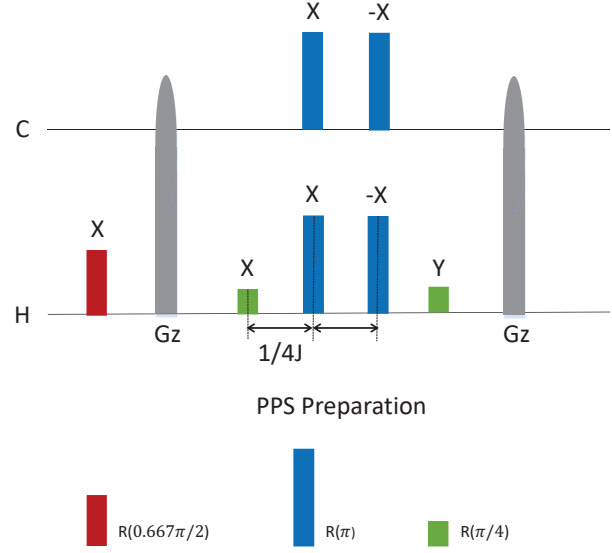}
\caption{ NMR sequences to realize pseudo-pure state. Gz means a $z$-gradient pulse which is used to cancel the polarization in $xy$ plane.}\label{fig:PPS}
\end{center}
\end{figure}

\subsection{Preparation of the Pseudo-pure State}
The thermal equilibrium state for the two qubit system is
\begin{equation}
\rho_{\textrm{eq}}\approx\frac{1}{4}I+\epsilon(\gamma_H\sigma^z_1+\gamma_C\sigma^z_2),
\end{equation}
where $I$ is a $4\times4$ identity matrix,
\begin{equation}
\epsilon\approx10^{-5}
\end{equation}
describes the polarization, and $\gamma_H$ and $\gamma_C$ represent the gyromagnetic ratios of the $^{1}$H and $^{13}$C nuclei, respectively. The spatial average technique \cite{Cory98} is used to obtain a pseudo-pure state
\begin{equation}
\rho_{00}=\frac{1-\epsilon}{4}I+\epsilon|00\rangle\langle00|
\end{equation}
and the related pulse sequence is depicted in Fig.~\ref{fig:PPS}. Thus we only focus on the part $\vert00\rangle$ as the entire system behaves since the identity part does not influence the unitary operations or measurements in NMR experiments.

\subsection{Evolution of the Hamiltonian with Debye noise}
The total Hamiltonian for simulating photosynthetic EET in the NMR system is
\begin{align}
H(t)&=H_S+n_1(t)\sigma^z_1+n_2(t)\sigma^z_2, \\
n_1(t)&=\frac{\beta_1(t)+\beta_2(t)}{2}, \\
n_2(t)&=\frac{\beta_1(t)-\beta_2(t)}{2},
\end{align}
where $H_S$ is the system Hamiltonian, and $\beta_{i}(t)$ ($i=1,2$) are the time-dependent Debye noises. In experiments, the evolution can have $L$ discretized steps, and the evolution time is $t=L\Delta t$ with
\begin{eqnarray}
U(t)&=&\exp[-iH(t)t]\notag\\
&=&\prod_{i=1}^L U_i\notag\\
&=&\prod_{i=1}^L \exp[-iH_i\Delta t],
\end{eqnarray}
where $H_i$ is the time-independent Hamiltonian at point $t_i=i\Delta t$. Note $U(t)$ is calculated by the gradient ascent pulse engineering (GRAPE) method with 5 ms of each pulse to reduce the accumulated pulse errors in experiments.

\subsection{Measure the Probability Distribution of Four States}
Our goal now is to acquire probability distributions of four states $|00\rangle$, $|01\rangle$, $|10\rangle$, $|11\rangle$; namely, the four diagonal values of the final density matrix. The density matrices of the output states are reconstructed completely via quantum state tomography (QST) \cite{Lee02}. In the QST theory, the density matrix of the system can be estimated from ensemble averages of a set of observables. For the one-qubit system, the observable set is $\{\sigma_i\}$ ($i=0,1,2,3$), where $\sigma_0=I$ is the identity, $\sigma_1=X$, $\sigma_2=Y$, $\sigma_3=Z$ are the Pauli matrices. The NMR signal is
\begin{equation}
S(t)\propto[\langle X\rangle+i\langle Y\rangle]e^{i\omega t},
\end{equation}
which is oscillating at the frequency $\omega$ and $\langle X\rangle$ and $\langle Y\rangle$ are obtained in practice by Fourier transforming $S(t)$ and integrating the real and imaginary spectra, respectively. The signal becomes
\begin{equation}
S^Y(t)\propto[-\langle Z\rangle+i\langle Y\rangle]e^{i\omega t}
\end{equation}
after applying $\exp[-i\pi Y/4]$. The density operator of one-qubit can be estimated by
\begin{equation}
\rho=\frac{1}{2}I+\langle X\rangle\ X+\langle Y\rangle Y+\langle Z\rangle\ Z.
\end{equation}
For the two-qubit system, the observable set is $\{\sigma_i\otimes\sigma_j\}(i,j=0,1,2,3)$. In our experiments, the complete density matrix tomography is not necessary. All we need is to perform two experiments in which the reading-out pulses $\exp(-i\pi Y/4)\otimes I$ and $I\otimes \exp(-i\pi Y/4)$ are respectively implemented on the final states of $^{1}$H and $^{13}$C and the corresponding qubits that need to be observed are respectively $^{1}$H and $^{13}$C. Then the probability distribution of four states are obtained by the results of measuring $\langle ZI\rangle$, $\langle I Z\rangle$, $\langle ZZ\rangle$. Then the probability distribution of four states are respectively
\begin{align}
\rho_{00}=\frac{1}{4}I+\langle ZI\rangle+\langle IZ\rangle+\langle ZZ\rangle,\\
\rho_{01}=\frac{1}{4}I+\langle ZI\rangle-\langle IZ\rangle-\langle ZZ\rangle,\\
\rho_{10}=\frac{1}{4}I-\langle ZI\rangle+\langle IZ\rangle-\langle ZZ\rangle,\\
\rho_{11}=\frac{1}{4}I-\langle ZI\rangle-\langle IZ\rangle-\langle ZZ\rangle.
\end{align}

\section{Gradient Ascent Pulse Engineering (GRAPE) Algorithm}
Here we describe the GRAPE technique proposed by Glaser \textit{et al.} \cite{Khaneja05} which has been frequently used in NMR experiments. For an $n$-qubit NMR system, the total Hamiltonian contains the internal term
\begin{equation}
H_t=H_{\textrm{int}}+H_{\textrm{RF}},
\end{equation}
and the radio frequency (RF) term
\begin{equation}
H_{\textrm{RF}}=-\sum_{k=1}^n\gamma_k B_k\left[\cos(\omega_{\textrm{RF}}^k t+\phi_k)\sigma_x^k+\sin(\omega_{\textrm{RF}}^k t+\phi_k)\sigma_y^k\right],
\end{equation}
where $B_k$ and $\phi_k$ are the amplitude and phase of the control field on the $k$th nuclear spin. The goal of the GRAPE technique is to find the optimal parameters $B_k$ and $\phi_k$ of the RF field by iteration to control the designed evolution $U_T$ very close to desired target evolution $U_D$. Assuming that the total time of RF field is $T$, which is divided into $L$ discrete segments. The time of each segment is $\Delta t=T/L$ and the time propagator of the $j$th segment can be expressed as
\begin{equation}
U_j=\exp[-i\Delta t(H_{\textrm{int}}+\sum_k u_x^k(j)\sigma_x^k+\sum_k u_x^k(j)\sigma_x^k)].
\end{equation}
Thus, the total evolution is
\begin{equation}
U_T=U_NU_{N-1}\cdots U_2U_1.
\end{equation}
The fidelity to the target evolution $U_{T}$ can be expressed as
\begin{equation}
\mathcal{F}=\frac{1}{2^n}|\textrm{Tr}(U_D^{\dagger}U_T)|,
\end{equation}
which is also called the fitness function. The GRAPE algorithm considers the fidelity $\mathcal{F}$ as the extreme value optimization of the multi-function. We calculate the gradient function to first order,
\begin{eqnarray}
g_{x,y}^k(j)&=&\frac{\partial F}{\partial u_{x,y}^k(j)}\notag\\
&\approx&-\frac{2}{2^n}\textrm{Re}[U_D^{\dagger}U_N\cdots(-i\Delta t\sigma_{x,y}^k)U_m\cdots U_1].
\end{eqnarray}
The fitness functions can be increased in the gradient iteration,
\begin{equation}
u_{x,y}^k(j)\rightarrow u_{x,y}^k(j)+\varepsilon\cdot g_{x,y}^k(j),
\end{equation}
where $\varepsilon$ is a suitable and small step size. The GRAPE procedure starts from an initial guess input and evaluates the corresponding gradients $g_{x,y}^k(j)$ and then keeps iterating until the fitness function reaches the desired value.

\section{Ramsey Fringe Experiment}

\begin{figure}[!ht]
\begin{center}
\includegraphics[width=8.5 cm,angle=0]{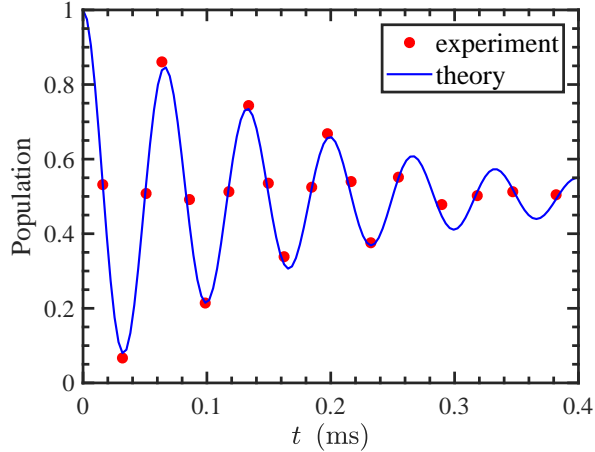}
\caption{The decay of the population of the $\vert0\rangle$ state. The theoretical simulation adopting the HEOM is shown by the blue curve, and the experimental data by the red dots. In our experiment, the parameters are $\varepsilon_{D}-\varepsilon_{A}=2\pi\times15~\rm{kHz}$, $\textrm{d}t=20~\rm{\mu s}$, $M=50$. The parameters of Debye noise are $\gamma_\textrm{NMR}=2\pi\times45~\rm{kHz}$, $\lambda_\textrm{NMR}=2\pi\times0.01~\rm{kHz}$, $T_\textrm{EET}=300~\rm{K}$, $\omega_{0}=2\pi\times5~\rm{Hz}$, $\omega_J=2\pi\times25~\rm{kHz}$.}
\label{noise}
\end{center}
\end{figure}

In Sec.~\ref{sec:Photosynthesis} and~\ref{sec:Ramsey}, we provide the deviation
for the Ramsey experiments in photosynthetic light harvesting and NMR respectively.
In this section, we will compare the data by these two different approaches, i.e.
the theoretical Ramsey fringes by HEOM and the experimental Ramsey fringes by NMR.

Ramsey experiments are often applied to observe the coherence of the qubits under the impact of noise. In our experiments, we perform the Ramsey spectroscopy for a single qubit subject to the Debye-Drude noise. In particular, the experiment is divided into three steps:
\begin{itemize}
 \item (1) Preparing the $\vert0\rangle$ state and rotating to the $xy$ plane by a $\pi/2$ pulse along the $x$-axis.
 \item (2) Free evolution while adding Debye-Drude noise by artificially injecting noise.
 \item (3) Applying a reverse $\pi/2$ pulse and measuring the population of the $\vert0\rangle$ state.
\end{itemize}

We observe that the decay time constant of the population of the $\vert0\rangle$ state is significantly reduced in the presence of the Debye noise, as shown in Fig.~\ref{noise}. Ramsey fringes are a fit to a cosine with a simple exponential decay envelope, see Eqs.~(\ref{P1}) and (\ref{P0}). The result by the HEOM theory (blue curve in Fig.~\ref{noise}) is consistent with that in experiment (red dots in Fig.~\ref{noise}). In other words, we demonstrate the equality of $\chi(t)$ and $\textrm{Re}[g(t)]$.

\section{Result of Numerical Simulation}
\label{sec:Result}

Before the experimental demonstration, we shall numerically demonstrate
the photosynthetic light harvesting can be exactly mimicked by the NMR
quantum simulation using the GRAPE algorithm.

In our numerical simulation, we used the following parameter, i.e.,
$\gamma_\textrm{NMR}=2\pi\times45~\textrm{kHz}$, $\lambda_\textrm{NMR}=2\pi\times0.01~\textrm{kHz}$, $T_\textrm{EET}=3\times10^4~\textrm{K}$, $T_\textrm{NMR}=5\times10^{-5}~\textrm{K}$. And we assume the other physical constants are one, that is, $\hbar=1.055\times10^{-34}~\rm{J\cdot s}$,
$k_{B}=1.381\times10^{-23}~\rm{J/K}$. The Hamiltonian for four-pigments in the single-excitation subspace is
\begin{equation}
H_\textrm{NMR}=2\pi\times\left(\begin{array}{cccc}650 & 6.3040 & 0.8059 & 0.2370 \\ 6.3040 & 645 & 6.5950 & 0.8059 \\ 0.8059 & 6.5950 & 615 & 6.3040 \\ 0.2370 & 0.8059 & 6.3040 & 610 \end{array}\right)\textrm{kHz}.
\end{equation}

\begin{figure}[!ht]
\begin{center}
\includegraphics[width=16 cm,angle=0]{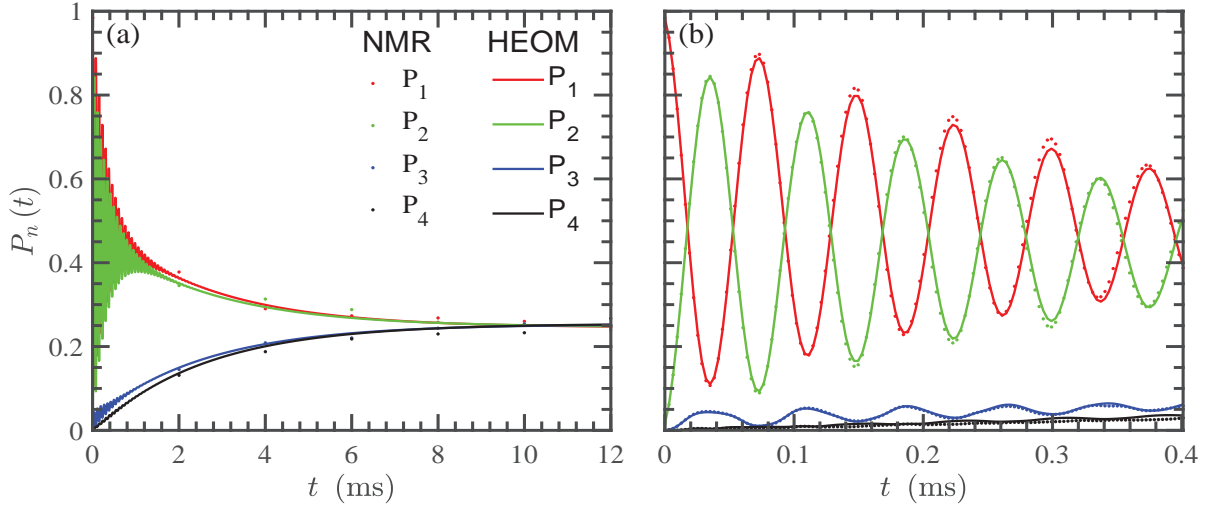}
\caption{The dots show the GRAPE numerical results and the lines show the HEOM results. The time interval of (a) is $0\sim12$ ms, and the time of (b) interval is
$0\sim0.5$ ms.}\label{figure12}
\end{center}
\end{figure}

\section{Computational Costs of NMR and HEOM}
\label{sec:CompuCost}

When the open quantum system consists of $N$ levels or sites, each coupled to an independent bath (so $N$ baths), and the correlation function of each bath contains $K$ exponentials, there are \cite{Ishizaki09,Shi09}
\begin{align}
\frac{(\mathcal{N}+KN)!}{\mathcal{N}!(KN)!} & \leq\frac{\sqrt{2\pi}(\mathcal{N}+KN)^{\mathcal{N}+KN+\frac{1}{2}}e^{-(\mathcal{N}+KN)}}{e\mathcal{N}^{\mathcal{N}+\frac{1}{2}}e^{-\mathcal{N}}eKN^{KN+\frac{1}{2}}e^{-KN}} \notag\\
 & =\sqrt{\frac{2\pi(\mathcal{N}+KN)}{\mathcal{N}KN}}\frac{(\mathcal{N}+KN)^{\mathcal{N}+KN}}\notag\\
 & =\sqrt{\frac{2\pi(\mathcal{N}+KN)}{e^{4}\mathcal{N}KN}}\left(1+\frac{KN}{\mathcal{N}}\right)^{\mathcal{N}}\left(1+\frac{\mathcal{N}}{KN}\right)^{KN}
\end{align}
density operators in a hierarchy with a cut-off of $\mathcal{N}$, where we have used the Stirling's formula \cite{Robbins55}. When the number of chlorophylls in the photosynthetic complex is very large,
e.g. 96 chlorophylls in PSI and about 300 chlorophylls in PSII,
and the form of the spectral density is complicated or the temperature is low,
\begin{align}
\lim_{K,N\rightarrow\infty}\frac{(\mathcal{N}+KN)!}{\mathcal{N}!(KN)!} & \leq\lim_{K,N\rightarrow\infty}\sqrt{\frac{2\pi(\mathcal{N}+KN)}{e^{4}\mathcal{N}KN}}\left(1+\frac{KN}{\mathcal{N}}\right)^{\mathcal{N}}\left(1+\frac{\mathcal{N}}{KN}\right)^{KN}\notag\\
 & =\sqrt{\frac{2\pi(\mathcal{N}+KN)}{e^{4}\mathcal{N}KN}}e^{\mathcal{N}+KN}.
\end{align}

On the other hand, the computational cost of GRAPE is \cite{Li17,Lu17}
\begin{equation}
4^{\log_2{N}}=N^2,
\end{equation}
where $N$ is the number of energy levels involved in the energy transfer.
Because the $N$-level photosynthetic light harvesting is simulated by $\log_2{N}$-qubit NMR,
the computational cost has the potential to be effectively reduced from exponential in $N$ by the HEOM to polynomial in $N$ by GRAPE.

\section{Effect of Number of Random Realizations and Error Analysis}

Errors are small and mainly caused by imperfections in the initial-ground-state preparation and GRAPE
pulses, which can be estimated by numerical simulations. The remaining errors may originate from, e.g., imperfections in the experimental quantum control, the static magnetic field, and the
spectral integrals.

\begin{figure*}[htbp]
\centering
\includegraphics[width=160mm,height=170mm]{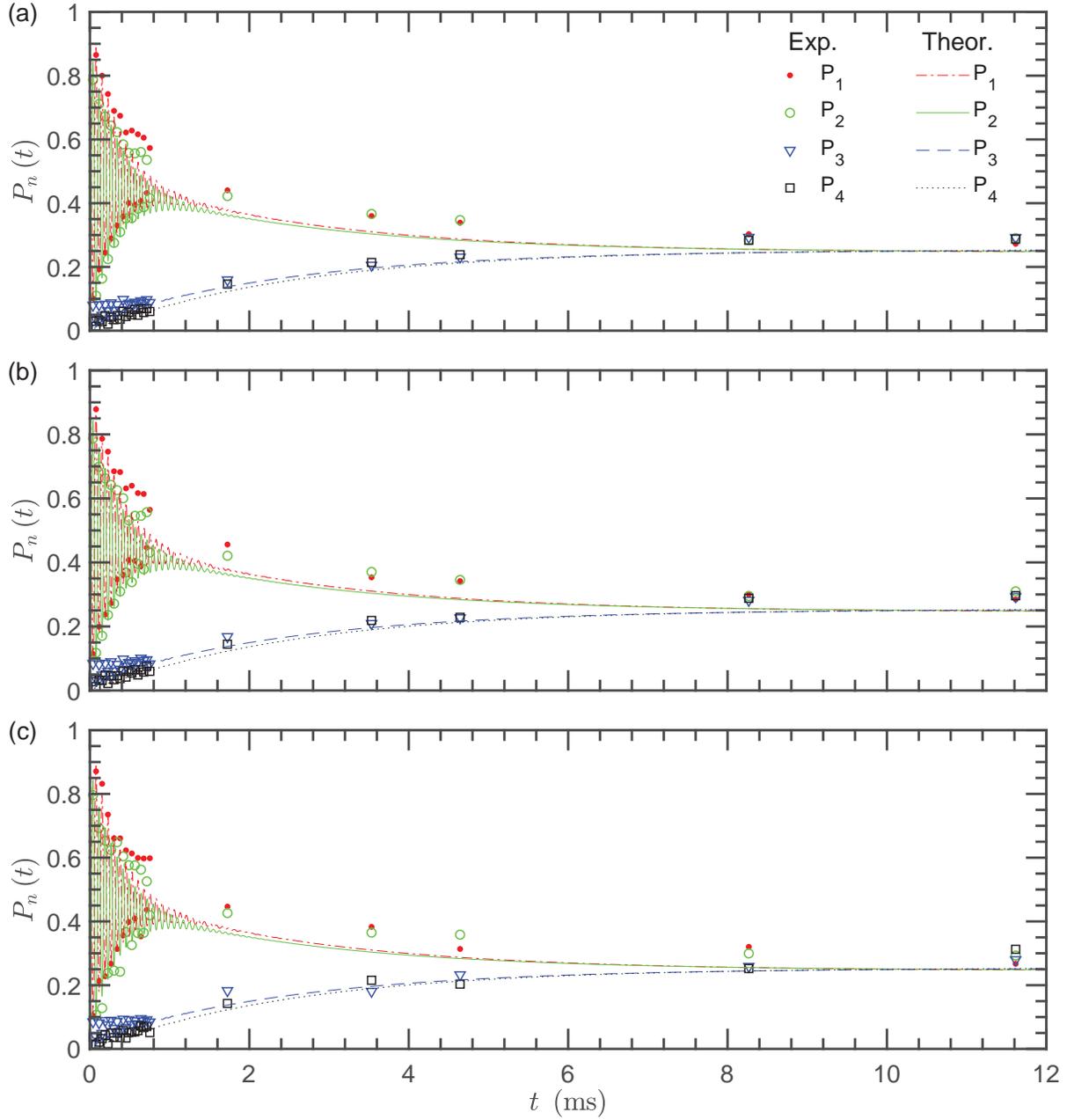}%
\caption{ Simulation of the energy transfer governed by $H_{\textrm{NMR}}$ for $M$ random realizations: (a) $M=50$, (b) $M=100$, and (c) $M=150$. The dots show the experimental data, and the curves are obtained from the numerical simulation using the HEOM. \label{fig:LongDynamics}}
\end{figure*}

\begin{figure*}[htbp]
\centering
\includegraphics[bb=0 110 600 350,width=160mm]{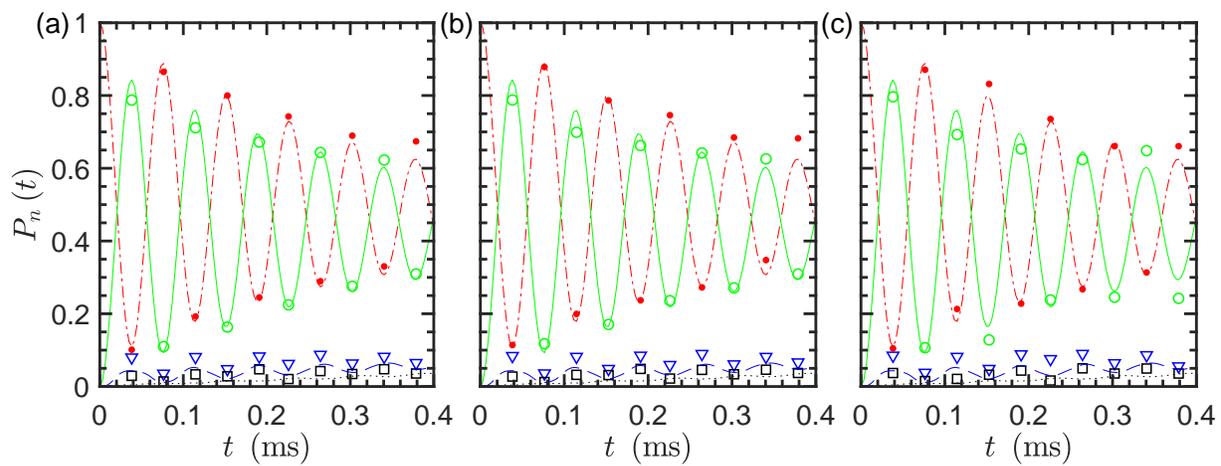}
\caption{Enlarged short-time regimes in Fig.~\ref{fig:LongDynamics}: (a) $M=50$; (b) $M=100$; (c) $M=150$. \label{fig:ShortDynamics}}
\end{figure*}

\begin{figure}[!ht]
\begin{center}
\includegraphics[width=16 cm,angle=0]{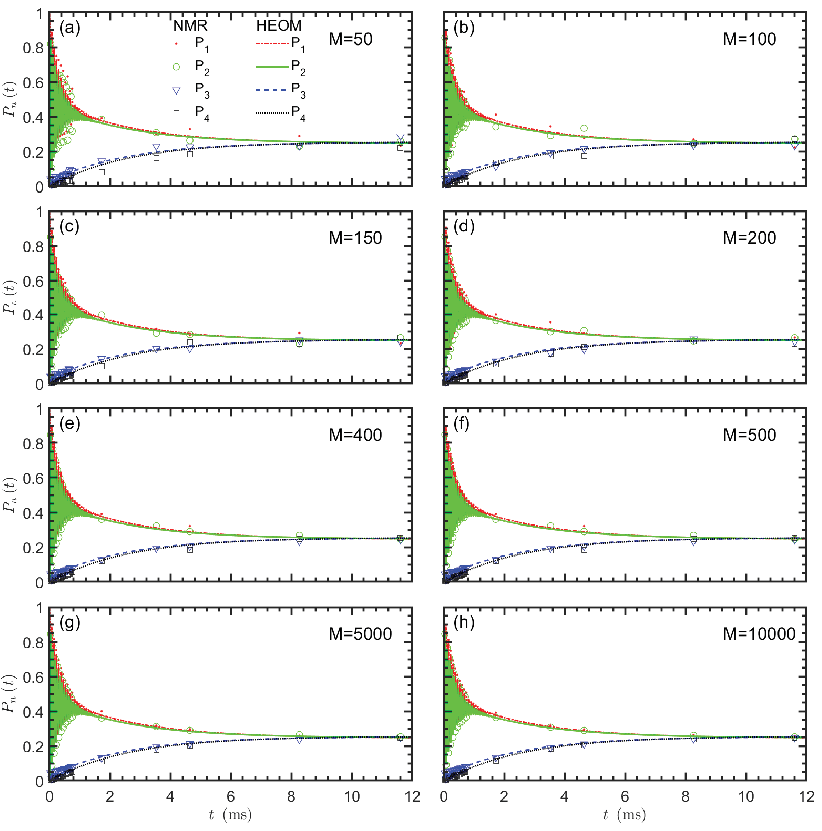}
\caption{The dots show the GRAPE numerical results and the lines show the HEOM results. The number of ensemble are $M=50,~100,~150,~200,~400,~500,~5000,~10000$ from (a) to (h).}\label{fig:EnsembleSize}
\end{center}
\end{figure}

As shown in Sec.~\ref{sec:Classical Noise}, in the derivation of quantum dynamics under the influence of noise, we have assumed that the average over random realizations is equivalent to the average over time. The assumption is valid only if the number of random realizations $M$ is in the infinite-$M$ limit. In order to verify this assumption, we experimentally investigate the effect of number of random realizations on the NMR simulation, as shown in Figs.~\ref{fig:LongDynamics},\ref{fig:ShortDynamics}. As $M$ increases from $M=50$ to $M=150$, the experimental simulation approaches closer and closer to the numerical simulation by the HEOM. In Fig.~\ref{fig:EnsembleSize},
 we further compare the numerical results by the GRAPE and the HEOM.
 As $M$ increases from $M=50$ to $M=10^4$, the difference between the results by the GRAPE and the HEOM reduces. When $M\geq500$, the difference is hardly noticeable. Therefore, it is justified to mimic the photosynthetic energy transfer by the NMR with random realizations.

\bibliographystyle{unsrt}

\begin{thebibliography}{99}

\bibitem{Cheng09} Cheng, Y.-C. \& Fleming, G. R.
    Dynamics of light harvesting in photosynthesis.
    \textit{Annu. Rev. Phys. Chem.} \textbf{60,} 241--262 (2009).

\bibitem{Romero17} Romero, E., Novoderezhkin, V. I. \& van Grondelle, R.
    Quantum design of photosynthesis for bio-inspired solar-energy conversion.
    \textit{Nature} \textbf{543,} 647--656 (2017).

\bibitem{Scholes17} Scholes, G. D. \textit{et al.}
    Using coherence to enhance function in chemical and biophysical systems.
    \textit{Nature} \textbf{543,} 647--656 (2017).

%

\bibitem{Lambert13} Lambert, N., Chen, Y.-N., Cheng, Y.-C., Chen, G.-Y. \& Nori, F.
    Quantum biology.
    \textit{Nature Phys.} \textbf{9,} 10--18 (2013).

\bibitem{Engel07} Engel, G. S. \textit{et al.}
    Evidence for wavelike energy transfer through quantum coherence in photosynthetic systems.
    \textit{Nature } \textbf{446,} 782--786 (2007).

\bibitem{Chin13} Chin, A. W. \textit{et al.}
    The role of non-equilibrium vibrational structures in electronic coherence and recoherence in pigment-protein complexes.
    \textit{Nature Phys.} \textbf{9,} 1--6 (2013).

\bibitem{Collini10} Collini, E. \textit{et al.}
    Coherently wired light-harvesting in photosynthetic marine algae at ambient temperature.
    \textit{Nature } \textbf{463,} 644--647 (2010).

\bibitem{Hildner13} Hildner, R., Brinks, D., Nieder, J. B., Cogdell, R. J. \& van Hulst, N. F.
    Quantum coherent energy transfer over varying pathways in single light-harvesting complexes.
    \textit{Science} \textbf{340,} 1448--1451 (2013).

\bibitem{Dong12} Dong, H., Xu, D. Z., Huang, J.-F. \& Sun, C.-P.
    Coherent excitation transfer via the dark-state channel in a bionic system.
    \textit{Light: Sci. Appl.} \textbf{1,} e2 (2012).

\bibitem{Tanimura06} Tanimura, Y.
    Stochastic Liouville, Langevin, Fokker-Planck, and master equation approaches to quantum dissipative systems.
    \textit{J. Phys. Soc. Jpn.} \textbf{75,} 082001 (2006).

\bibitem{Ishizaki09} Ishizaki, A. \& Fleming, G. R.
    Theoretical examination of quantum coherence in a photosynthetic system at physiological temperature.
    \textit{Proc. Natl. Acad. Sci. U.S.A.} \textbf{106,} 17255--17260 (2009).

\bibitem{Ai13} Ai, Q., Yen, T.-C., Jin, B.-Y. \& Cheng, Y.-C.
    Clustered geometries exploiting quantum coherence effects for efficient energy transfer in light harvesting.
    \textit{J. Phys. Chem. Lett.} \textbf{4,} 2577--2584 (2013).

\bibitem{Buluta09} Buluta, I. \& Nori, F.
    Quantum simulators.
    \textit{Science} \textbf{326,} 108--111 (2009).

\bibitem{Georgescu14} Georgescu, I., Ashhab, S. \& Nori, F.
    Quantum simulation.
    \textit{Rev. Mod. Phys.} \textbf{86,} 153--185 (2014).

\bibitem{Buluta11} Buluta, I., Ashhab, S. \& Nori, F.
    Natural and artificial atoms for quantum computation.
    \textit{Rep. Prog. Phys.} \textbf{74,} 104401 (2011).

\bibitem{Soare14} Soare, A. \textit{et al.}
    Experimental noise filtering by quantum control.
    \textit{Nature Phys.} \textbf{10,} 825--829 (2014).

\bibitem{Zhen16} Zhen, X.-L., Zhang, F.-H., Feng, G. R., Li, H.
\& Long, G.-L.
    Optimal experimental dynamical decoupling of both longitudinal and transverse relaxations.
    \textit{Phys. Rev. A} \textbf{93,} 022304 (2016).

\bibitem{Luo17} Luo, Z. H. \textit{et al.}
    Experimentally probing topological order and its
    breakdown through modular matrices.
    \textit{Nature Phys.} in press (2017) doi:10.1038/nphys4281.

\bibitem{Feng13} Feng, G. R., Xu, G. F. \& Long, G. L.
    Experimental realization of nonadiabatic holonomic quantum computation.
    \textit{Phys. Rev. Lett.} \textbf{110,} 190501 (2013).

\bibitem{Chuang98} Chuang, I. L., Vandersypen, L. M. K., Zhou, X. L.,
 Leung, D. W. \& Lloyd, S.
    Experimental realization of a quantum algorithm.
    \textit{Nature} \textbf{393,} 143--146 (1998).

\bibitem{Khaneja05} Khaneja, N., Reiss, T., Kehlet, C. \& Schulte-Herbr\"{u}ggen, T.
    Optimal control of coupled spin dynamics: Design of NMR pulse sequences by gradient ascent algorithms.
    \textit{J. Mag. Res.} \textbf{172,} 296--305 (2005).

\bibitem{Zhang11} Zhang, J. F., Wei, T.-C. \& Laflamme, R.
    Experimental quantum simulation of entanglement in many-body systems.
    \textit{Phys. Rev. Lett.} \textbf{107,} 010501 (2011).

\bibitem{Peng14} Peng, X. H. \textit{et al.}
    Experimental implementation of adiabatic passage between different topological orders.
    \textit{Phys. Rev. Lett.} \textbf{113,} 080404 (2014).

\bibitem{Li17} Li, J., Yang, X. D., Peng, X. H., \& Sun, C.-P.
    Hybrid quantum-classical approach to quantum optimal control.
    \textit{Phys. Rev. Lett.} \textbf{118,} 150503 (2017).

\bibitem{Lu17} Lu, D. W. \textit{et al.}
    Quantum state tomography via reduced density matrices.
    \textit{npj Quantum Inf.} \textbf{3,} 45 (2017).

\bibitem{Potonik17} Pot\v{o}nik, A. \textit{et al.}
    Studying light-harvesting models with superconducting circuits.
    \textit{arXiv:} 1710.07466 (2017).

\bibitem{Rey13} del Rey, M., Chin, A. W., Huelga, S. F. \& Plenio, M. B.
    Exploiting structured environments for efficient energy transfer: The phonon antenna mechanism.
    \textit{J. Phys. Chem. Lett.} \textbf{4,} 903--907 (2013).

\bibitem{Gershenfeld97}  Gershenfeld, N. A. \& Chuang, I. L.
    Bulk spin-resonance quantum computation.
    \textit{Science} \textbf{275,} 350--356 (1997).

\bibitem{Cory97} Cory, D. G., Fahmy, A. F. \& Havel, T. F.
    Ensemble quantum computing by NMR spectroscopy.
    \textit{Proc. Natl. Acad. Sci. U.S.A.} \textbf{94,} 1634--1639 (1997).

\bibitem{Xin17} Xin, T. \textit{et al.}
    Quantum state tomography via reduced density matrices.
    \textit{Phys. Rev. Lett.} \textbf{118,} 020401 (2017).

%


\end{thebibliography}

\begin{thebibliography}{99}

\bibitem{Cheng09} Cheng Y.-C. \& Fleming, G. R.
    Dynamics of light harvesting in photosynthesis.
    \textit{Annu. Rev. Phys. Chem.} \textbf{60}, 241--262 (2009).

\bibitem{Lambert13} Lambert, N., Chen, Y.-N., Cheng, Y.-C., Chen, G.-Y. \& Nori, F.
    Quantum biology.
    \textit{Nature Phys.} \textbf{9,} 10--18 (2013).

\bibitem{Green13} Green, T. J., Sastrawan, J., Uys, H. \& Biercuk, M. J.
    Arbitrary quantum control of qubits in the presence of universal noise.
    \textit{New J. Phys.} \textbf{15}, 095004 (2013).

\bibitem{Soare14} Soare, A. \textit{et al.}
    Experimental bath engineering for quantitative studies of quantum control.
    \textit{Phys. Rev. A} \textbf{89}, 042329 (2014).

\bibitem{Zhen16} Zhen, X.-L., Zhang, F.-H., Feng, G.-R., Li, H. \& Long, G.-L.
    Optimal experimental dynamical decoupling of both longitudinal and tranverse relaxations.
    \textit{Phys. Rev. A} \textbf{93}, 022304 (2016).

\bibitem{Sakurai11} Sakurai, J. J.
    \textit{Modern Quantum Mechanics} (Pearson Education, 2011).

\bibitem{Tanimura06} Tanimura, Y.
    Stochastic Liouville, Langevin, Fokker¨CPlanck, and master equation approaches to quantum dissipative systems.
    \textit{J. Phys. Soc. Jpn.} \textbf{75,} 082001 (2006).

\bibitem{Ishizaki091} Ishizaki, A. \& Fleming, G. R.
    Unifield treatment of quantum coherent and incoherent hopping dynamics in electronic energy transfer: Reduced hierarchy equation approach.
    \textit{J. Chem. Phys.} \textbf{130}, 234111 (2009).

\bibitem{Ishizaki092} Ishizaki, A. \& Fleming, G. R.
    Theoretical examination of quantum coherence in a photosynthetic system at physiological temperature.
    \textit{Proc. Natl. Acad. Sci. U.S.A.} \textbf{106}, 17255--17260 (2009).

\bibitem{Novoderezhkin10} Novoderezhkin, V. I. \& van Grondelle, R.
    Physical origins and models of energy transfer in photosynthetic light-harvesting.
    \textit{Phys. Chem. Chem. Phys.} \textbf{12}, 7352--7265 (2010).

\bibitem{Grabert88} Grabert, H., Schramm, P. \& Ingold, G.-L.
    Quantum Brownian motion: The functional integral approach.
    \textit{Phys. Rep.} \textbf{168}, 115--207 (1988).

\bibitem{Cory98} Cory, D. G., Price, M. D. \& Havel, T. F.
    Nuclear magnetic resonance spectroscopy: An experimentally accessible paradigm for quantum computing.
    \textit{Phys. D} \textbf{120}, 82--101 (1998).

\bibitem{Lee02} Lee, J.-S.
    The quantum state tomography on an NMR system,
    \textit{Phys. Lett. A} \textbf{305}, 349--353 (2002)

\bibitem{Khaneja05} Khaneja, N., Reiss, T., Kehlet, C., Schulte-Herbr$\ddot{u}$ggen, T. \& Glaser, S. J.
    Optimal control of coupled spin dynamics: design of NMR pulse sequences by gradient ascent algorithms.
    \textit{J. Magn. Reson.} \textbf{172}, 296--305(2005).

\bibitem{Blanes09} Blanes, S., Casas, F., Oteo, J. A. \& Ros, J.
    The Magnus expansion and some of its applications.
    \textit{Phys. Rep.} \textbf{470}, 151--238 (2009).

\bibitem{Goodman15} Goodman, J. W.
    \textit{Statistical Optics} 2nd Ed. (Wiley, Hoboken, NJ, 2015).

\bibitem{Miller12} Miller, S. L. \& Childers, D.
    \textit{Probability and Random Processes with Applications to Signal Processing and Communications} (Academic, Boston, MA, 2012).

\bibitem{Mukamel95} Mukamel, S.
    \textit{Principle of Nonlinear Optical Spectroscopy} (Oxford University Press, New York, 1995).

\bibitem{Ishizaki09} Ishizaki, A. \& Fleming, G. R.
    Theoretical examination of quantum coherence in a photosynthetic system at physiological temperature.
    \textit{Proc. Natl. Acad. Sci. U.S.A.} \textbf{106,} 17255--17260 (2009).

\bibitem{Robbins55} Robbins, H.
    A remark on Stirling's formula.
    \textit{Am. Math. Mon.} \textbf{62}, 26--29 (1955).

\bibitem{Li17} Li, J., Yang, X. D., Peng, X. H. \& Sun, C.-P.
    Hybrid quantum-classical approach to quantum optimal control.
    \textit{Phys. Rev. Lett.} \textbf{118,} 150503 (2017).

\bibitem{Lu17} Lu, D. W. \textit{et al.}
    Quantum state tomography via reduced density matrices.
    \textit{npj Quantum Inf.} \textbf{3,} 45 (2017).

\bibitem{Adolphs06} Adolphs, J. \& Renger, T.
    How proteins trigger excitation energy transfer in the FMO complex of green sulfur bacteria.
    \textit{Biophys. J.} \textbf{91,} 2778--2797 (2006).

\bibitem{Shi09} Shi, Q., Chen, L., Nan, G., Xu, R.-X. \& Yan, Y. J.
    Efficient hierarchical Liouville space propagator to quantum dissipative dynamics.
    \textit{J. Chem. Phys.} \textbf{130,} 084105 (2009).


\end{thebibliography}

\end{document}